\RequirePackage{fix-cm}
\documentclass[smallextended]{svjour3}       
\smartqed  
\usepackage{graphicx}
\usepackage{mathptmx}      
\usepackage{latexsym}
\usepackage{natbib}
\def\apjl{{Ap.J.(Lett)}}

\def\kms{km s$^{-1}$}
\def\sqd{{deg$^{2}$}}
\def\arcmin{$^{\prime}$}
\def\arcsec{$^{\prime\prime}$}
\def\LCDM{$\Lambda$CDM}
\def\msun{M$_\odot$}

\def\lsim{\raise0.3ex\hbox{$<$}\kern-0.75em{\lower0.65ex\hbox{$\sim$}}}
\def\gsim{\raise0.3ex\hbox{$>$}\kern-0.75em{\lower0.65ex\hbox{$\sim$}}}

\def\be{\begin{equation}}
\def\ee{\end{equation}}
\begin{document}

\title{Extragalactic HI Surveys
}


\author{Riccardo Giovanelli         \and
        Martha P. Haynes 
}


\institute{R. Giovanelli \at
              302 Space Science Building, Cornell University, Ithaca NY 14853 USA \\
              Tel.: +001-607-2556505\\
              \email{riccardo@astro.cornell.edu}           
           \and
           M.P. Haynes \at
              530 Space Science Building, Cornell University, Ithaca NY 14853 USA \\
              Tel.: +001-607-2550610\\
              \email{haynes@astro.cornell.edu}           
}

\date{Received: date / Accepted: date}

\maketitle

\begin{abstract}
We review the results of HI line surveys of extragalactic sources in the local Universe.
In the last two decades major efforts have been made in establishing on firm statistical
grounds the properties of the HI source population, the two most prominent being the HI Parkes 
All Sky Survey (HIPASS) and the Arecibo Legacy Fast ALFA survey (ALFALFA). We review the
choices of technical parameters in the design and optimization of spectro-photometric ``blind''
HI surveys, which for the first time produced extensive HI-selected data sets. Particular
attention is given to the relationship between optical and HI populations, the differences
in their clustering properties and the importance of HI-selected samples in contributing
to the understanding of apparent conflicts between observation and theory on the abundance
of low mass halos. The last section 
of this paper provides an overview of currently ongoing and planned surveys which will explore
the cosmic evolution of properties of the HI population. 

\end{abstract}

\section{Introduction}
\label{intro}

A number of important HI line surveys have been carried out in the last
decade, bringing new insights in particular to the properties of the 
extragalactic population at $z\simeq 0$. Those surveys have been made
possible by the advent of focal plane receiver arrays and telescope
upgrades, which have increased the survey speeds by an order of
magnitude with respect to what was possible before their inception.
While large surveys of HI in galaxies had been carried out in the
1980s and 1990s, those were targeted on optically selected samples
generally motivated by goals other than the understanding of the
characteristics of the HI source population. For example, large samples
of disk galaxies have been observed to map the large-scale structures
in the local Universe, or to measure the cosmological peculiar velocity 
field, or study the environment in clusters, but in those studies HI
sources were tools, the support riders in the {\it peloton}, rather
than the cycling team leader. We learned much about the properties
that made an HI source a good cosmic lamp post, but we did not know
the shape of the cosmic HI mass function. The first ``blind'' HI
surveys, such as AHISS and ADBS, started bringing to our attention the
characteristics of the dwarf galaxy population, and the following larger
surveys confirmed the fact that the most abundant ``island universes''
are in fact low mass systems barely able to hold on to a fraction of 
their baryons and, even more extreme, far less able to convert those
baryons into stars. Some toyed with the idea that these systems would
be so abundant that they would fill the gap between the counts of low 
mass galaxies and that of low mass halos predicted by theory,
known and not satisfactorily explained since the onset of the \LCDM ~paradigm.
The now solid determination of the HI mass function demonstrates that
idea to be wrong. However, the new surveys are suggesting interesting
avenues that may lead to a solution of the ``problem'', so we devote
a fair fraction of this review to the investigation of these developments.

We start Section \ref{sec:review} 
by visiting a set of scaling relations, useful
in the planning and design of a survey. We close that section with an
illustration of the power of a blind survey data set in reaching very
deep in signal extraction, by careful application of stacking techniques.
In Section \ref{sec:properties} we review the properties of the HI source
population, from the clustering characteristics to the HI mass function,
the relationships between gas content and properties in spectral domains
other than HI, to the HI view of star formation laws, closing with the
baryonic mass function and the assessment of the baryon deficit of small
mass halos.  Section \ref{dark} ~is a show-and-tell interlude, in which
we describe the extreme properties of a few sources in the ALFALFA
catalog. Section \ref{tbtf} explores the conflicts between observations and 
\LCDM ~predictions
arising from the use of the abundance matching technique. The last section
is dedicated to a highlights report of proposed (and some already started) new
surveys. Sky positions are for the epoch 2000.0 and distances are derived
for $H_\circ=70$ \kms ~Mpc$^{-1}$, unless otherwise specified.

\section{Review and Status of Large HI Surveys}
\label{sec:review}

\subsection{Survey Design Tools}
\label{sec:criteria}

\begin{figure}[t]
  \includegraphics[width=1.0\textwidth]{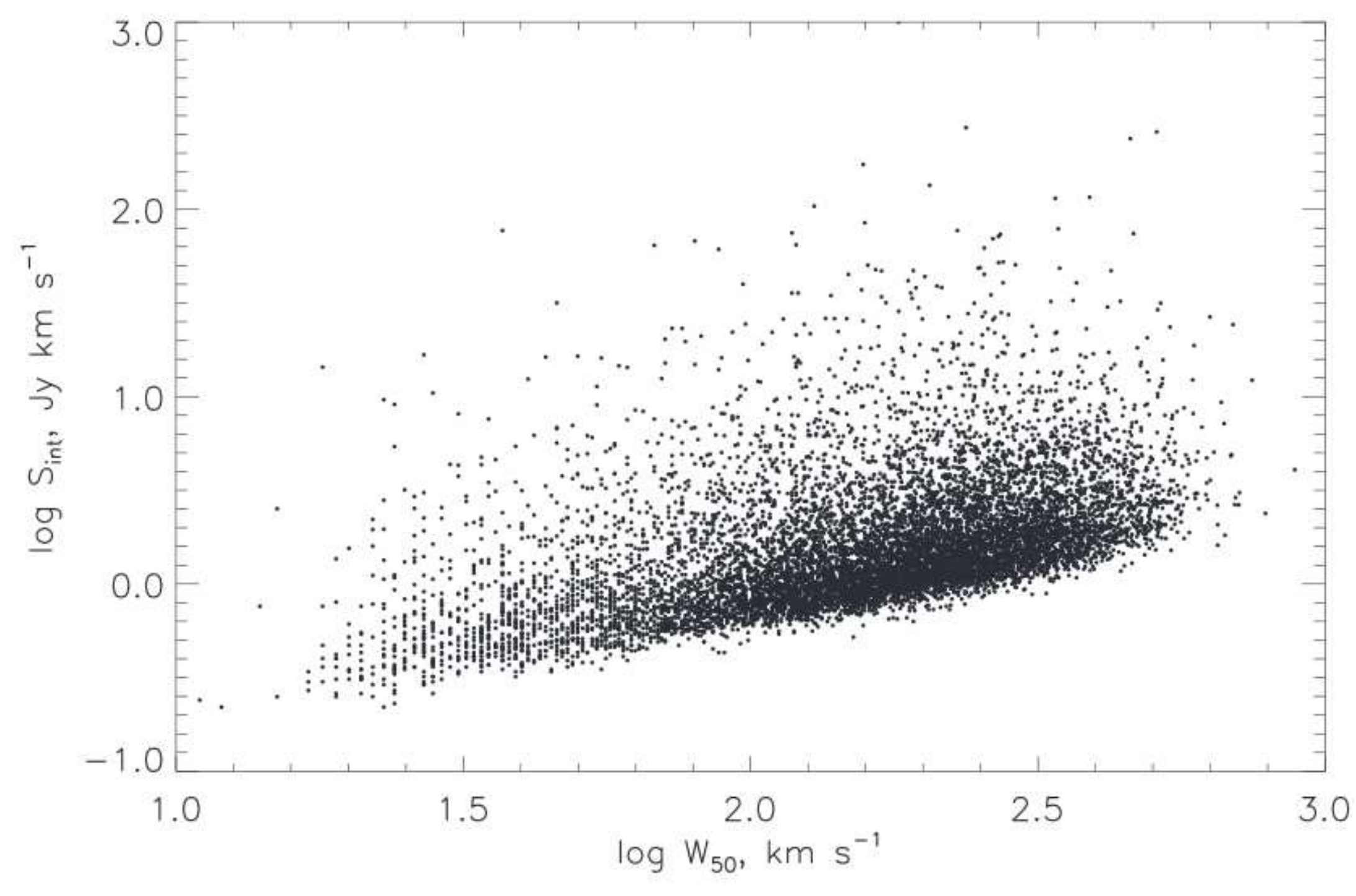}
\caption{\footnotesize {Log of the HI flux vs. velocity width of HI sources 
detected by ALFALFA. $\gamma$ refers to the slope of the lower boundary
of the data points, which steepens for $W50>200$ \kms.
}}
\centering
\label{fig:fluxlim}       
\end{figure}

The sensitivity of a radio telescope depends primarily on (a) the system's ``gain'' $G$,
which to first order is the area of its primary mirror; (b) the noise contributed
by the electronics, ground, atmospheric and cosmic sources of radiation, which is 
measured by the system temperature $T_{sys}$; and (c) the frequency bandwidth $\Delta\nu$
over which radiation is collected. This is summarized by the radiometer equation
\be 
\sigma_{rms} = (T_{sys}/ G)(2 ~\Delta\nu ~t_{int}f_x)^{-1/2}
\ee
where $\sigma_{rms}$ is the rms noise of a spectrum with the telescope pointing at
blank sky, $t_{int}$ is the integration time, $f_x$, of
order unity, accounts for source angular extent, observational switching technique, 
bandpass subtraction mode, spectrometer clipping losses, etc., and the factor of 2 
accounts for the addition of two independent polarization channels. In the computation 
of sensitivity, $\Delta\nu$ will refer to the spectral resolution needed to detect and 
resolve the narrowest signal the survey will be sensitive to; in the estimate of survey 
speed, $\Delta\nu$ will be the instantaneous bandpass of the receiver system. Most of the 
HI in a galaxy is found in a warm, neutral, optically thin thermal phase of several $10^3$ K and the
solid angle subtended by the cold neutral phase of the interstellar medium is small, so the 
mass of a spatially unresolved, optically thin source at the luminosity distance $D_{Mpc}$, 
expressed in Mpc, is
\be
{M_{HI}\over M_\odot}={2.356\times 10^5 ~D_{Mpc}^2 \over 1+z} \int S(V) dV,
\ee
where $S(V)$ is the line profile in Jy, integrated over the Doppler velocity $V$ 
in \kms. Approximating $\int S(V)dV \simeq S_{peak} W_{50}$, the product of peak flux 
times the line width at the 50\% level, the signal-to-noise ratio 
($S_{peak}/\sigma_{rms}$) can be used to infer the scaling of the
minimum detectable HI mass at the distance $D_{Mpc}$
\be
M_{HI} \propto (T_{sys}/G) ~D_{Mpc}^2 W_{50}^{-\gamma} ~t_{int}^{-1/2}
\ee
Figure \ref{fig:fluxlim}
shows the flux distribution of sources detected by ALFALFA, as a function of their 
line widths. The exponent $\gamma$ is -1 for $W_{50} < 200$ \kms, and it changes 
progressively to $\gamma \simeq -2$
for $W_{50} > 200$ \kms, due to the impact of standing waves on the spectral baselines.

A blind, spectro-photometric HI survey does not deliver a flux-limited source 
sample, but rather one in which the sensitivity limit is
a function of the spectral linewidth of the source.
\begin{figure}
  \includegraphics[width=1.0\textwidth]{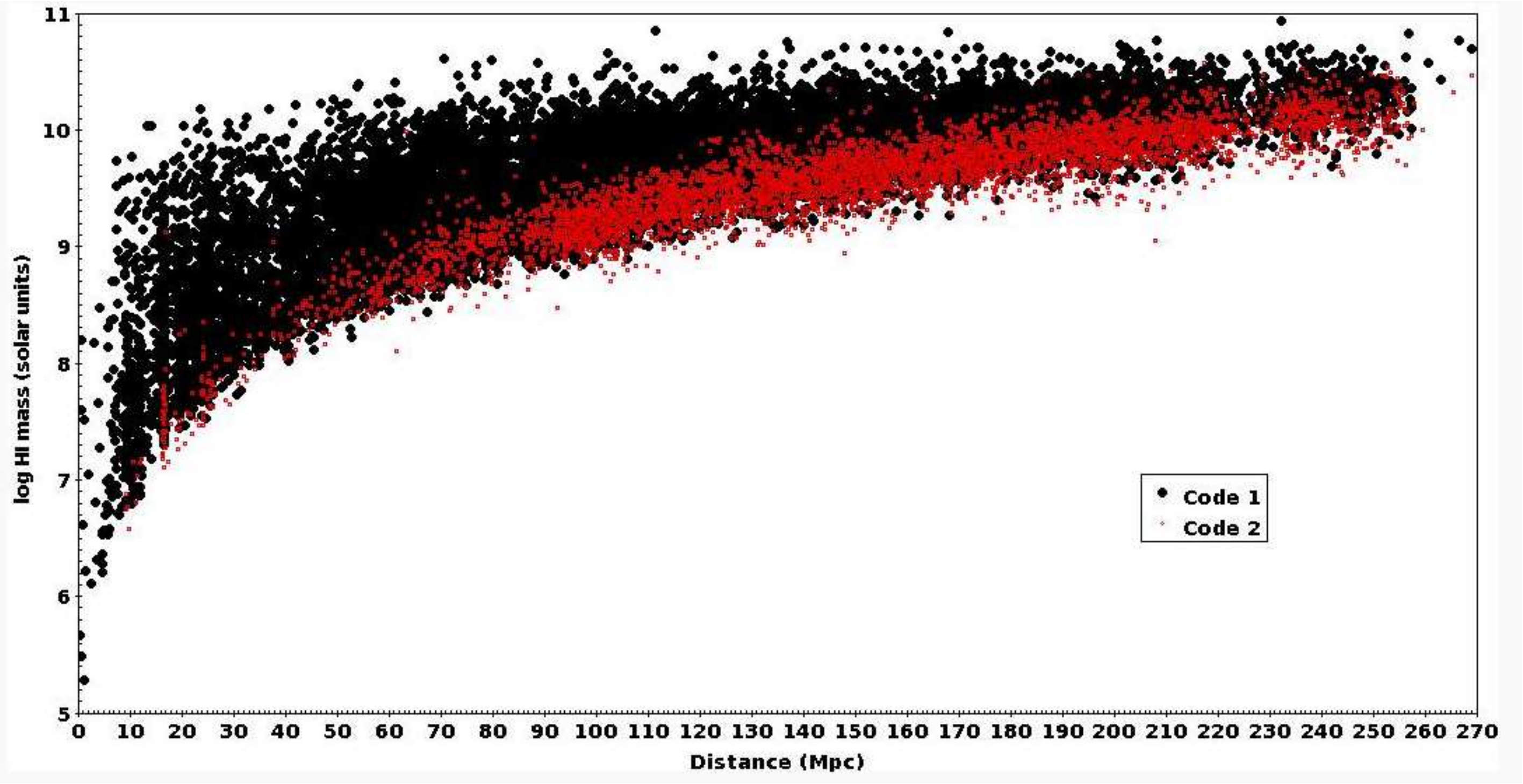}
\caption{\footnotesize {Log of the HI mass distribution vs. distance of HI sources 
detected by ALFALFA. Black symbols identify sources of high signal-to-noise - greater
than 6.5 -, while red symbols refer to sources of signal-to-noise between 4.5 and 6.5,
corroborated by having an a priori known redshift which matches the ALFALFA detection.
A gap in the data near $D_{Mpc}\simeq 230$ is produced by radio fequency interference
(RFI) originating in the San Juan airport radar; a lesser dip near $D_{Mpc}\simeq 85$ 
is due to transmissions associated with the Global Positioning System (GPS), as we
discuss in Section \ref{rfi}.
}}
\label{fig:Spaen}       
\end{figure}
In the case of the ALFALFA survey, with an average integration time $t_{int}\simeq 48$
seconds per beam solid angle, an HI source with $W_{50}=25$ \kms ~and $M_{HI}=10^6$ \msun
~can be detected to a distance of 6.5 Mpc, and a source of HI mass $10^7$ \msun ~to a distance
of 20 Mpc, as illustrated in Figure \ref{fig:Spaen}.
 
A useful scaling relation regards the minimum integration time required to detect 
a source of HI mass $M_{HI}$, line width $W_{50}$ \kms, at the distance $D_{Mpc}$ 
\be 
t_{int} \propto (T_{sys}/G)^2 ~M_{HI}^{-2} ~D_{Mpc}^4 ~W_{50}^{-2\gamma}
\ee
i.e. the depth of a survey of given $T_{sys}/G$ increases only as $t_{int}^{1/4}$.
For a given HI mass, the volume sampled by a survey with a telescope of given $T_{sys}/G$
is 
\be 
V_{survey}(M_{HI}) = \Omega_{survey} ~D^3_{max}/3 \propto \Omega_{survey} t_{int}^{3/4}
\ee
where $\Omega_{survey}$ is the solid angle covered by the survey and $D_{max}$ is the maximum distance
at which the mass $M_{HI}$ can be detected. That volume is proportional to the number of sources detected
by the survey, so the time necessary to complete it is 
$t_{survey}\propto (\Omega_{survey}/\Omega_b) t_{int}$, where $\Omega_b$ is the telescope beam (or
beams, for arrays), thus
\be
t_{survey} \propto V_{survey}~ t_{int}^{1/4}
\ee
i.e. for a fixed $\Omega_{survey}$, the time required to complete it increases as $t_{int}^{1/4}$.
Once $t_{int}$ is large enough for a desired $M_{HI}$ to be detectable, it is more advantageous
to maximize $\Omega_{survey}$ than to increase the survey depth by increasing $t_{int}$. For example, 
in the design of the ALFALFA survey it was desired that the survey sample median distance
correspond to a fair cosmological distance of $\simeq$100 Mpc, the choice of which determined
$t_{int}$. After that, the survey volume was obtained by expanding $\Omega_{survey}$.

In the case of blind surveys carried out with a single dish, such as HIPASS and ALFALFA, the
vast majority of detected sources have angular size much smaller than the telescope beam.
The photometric parameter used for their characterization is then the line flux integral,
information on the morphology and spatial extent of the source being lost. A lower limit 
for the averaged HI column density of the source can be obtained, by dividing the HI
mass by the areal extent subtended by the beam. For a telescope with an elliptical beam, 
such as the case of the Arecibo antenna with its 7-horn feed array ("ALFA"), with major and minor axes $\theta_1$ and 
$\theta_2$ (FWHP, in arcminutes):

\begin{equation}
\bar{N}_{HI} = { {2.34 \times 10^{20}} \over {\theta_1 \theta_2}} (1 + z)^4
     \int s(\nu) d\nu ~~~(cm^{-2})  
\label{eq:aobeam}
\end{equation}

Large single dishes with focal plane detector arrays, such as the Arecibo and Parkes
telescopes, are well suited to carry out wide field, blind HI extragalactic surveys,
such as ALFALFA and HIPASSS. However, follow-up synthesis imaging is necessary 
in order to obtain direct information on dynamical parameters, such as
the mass contained within the HI radius of the source and the shape of the rotation curve.
Synthesis imaging can provide maps of HI column densities, but can wholly miss flux from 
diffuse regions extending over solid angles exceeding that of the synthesized beam. A
combination of both types of instrument is thus the most economical and scientifically rewarding:
the single dish survey is most efficient in delivering large samples, characterizing
the statistical properties of the HI source population, and discovering the most extremes
member of such population; the synthesis arrays allow the investigation of the structure,
dynamics and physical circumstances of selected sources. 

\subsection{Types of Surveys}

Before the 1990s, large extragalactic HI surveys were carried out, seeking detection 
of optically selected galaxy samples. For these surveys, the task of understanding 
the properties of the HI source population was secondary to that of using the HI 
source samples as tools towards the achievement of other scientific goals, such as 
the impact of environment on gas bearing objects in clusters, the determination of 
the parameters of the cosmological distance scale, the investigation of the topology 
of the large scale distribution of galaxies and mapping the deviations from smooth 
Hubble flow produced by large scale density inhomogeneities.
The studies with the largest numbers $N_{det}$ of detections  were obtained with 
the Arecibo ($N_{det} \sim 10^4$),
Nan\c{c}ay, Parkes and Green Bank (a few $10^3$ each) telescopes. The most conspicuous 
bias in the selection of sample candidates for these studies was a preference for
high optical luminosity systems, morphologically classified as disks, which tended to
neglect field dwarfs.
An interesting statistic illustrating this bias is that more than 2/3 of detections
by ALFALFA correspond to galaxies never observed before in the HI line.
Results of these studies are documented in reviews and data releases such as Haynes 
et al. (1984), Aaronson et al. (1982), Fisher \& Tully (1981), Giovanelli \& Haynes (1991),
Roberts \& Haynes (1994), Bottinelli et al. (1990), Haynes et al. (1999), Mathewson \&
Ford (1996). These surveys will not be reviewed further.

The first strictly blind and moderately blind extragalactic HI surveys were made
at Arecibo in the late 1990's, during the period in which the telescope was being
upgraded. Two surveys (AHISS: Zwaan et al. 1997; ADBS: Rosenberg \& Schneider 2000)
provided the first indications of the statistical importance of the field dwarf
galaxies, whose baryonic component is dominated by atomic gas rather than stars.
HIPASS (Meyer et al. 2004) was the first truly wide-field, extragalactic HI survey,
covering 3/4 of the whole sky with its 13-element focal plane detector array.

With its large collecting aperture, the 53 year old 305m telescope at Arecibo 
remains the most sensitive HI detector. During its first 30 years of operation,
its use as a blind survey machine was limited due to the fact that the correction 
of the spherical aberration of its primary mirror was achieved by means of line feeds. 
These are waveguide contraptions with very narrow instantaneous bandwidth and lacking 
a usable, extended field of view, which constrained to single pixel observations.
Completion of the Gregorian Upgrade in the late 1990s expanded the field of view,
making possible the installation of receiver arrays, broadening the bandwidth to 300 MHz;
sensitivity was further increased by improving the illuminated area of the primary 
mirror and installing a 1-km long screen, peripheral to the primary, that reduced 
ground radiation pick-up and improved performance at high zenith angles. Operating 
in L-band, the 7-beam ALFA (Arecibo L-band Feed Array) array became available 
in 2005. Several HI surveys with ambitious goals were then started. Of particular 
interest were the four extragalactic HI surveys: Arecibo Legacy Fast ALFA (ALFALFA, 
Giovanelli et al. 2005), Arecibo Galactic Environment Survey (AGES, Auld et al. 2006), 
ALFA Zone of Avoidance Deep Survey (ZOA, Henning et al. 2008) and Arecibo Ultra Deep 
Survey (AUDS, Freudling et al. 2011). 

A comparison of major blind surveys of the last
two decades is shown in Table \ref{tab:blindtab}, which illustrates angular resolution, 
survey solid angle, spectral resolution, noise figure, median redshift and number of 
detections, respectively from columns 2 to 7. The noise figure is expressed for a 
common value of spectral resolution of 18 \kms, although several surveys have observed
with much higher spectral resolution. The AUDS and ZOA surveys have reported on
preliminary results (Henning et al. 2010; McIntyre et al. 2015;  
Hoppmann et al. 2015) and have plans for more extensive work. The AGES team
is accumulating data over a number of specially targeted regions spanning a
range of environments, from isolated systems to the Virgo and A1367 clusters. 
Preliminary AGES results (Davies et al. 2011) suggest a steeper
mass function ($\alpha \sim1.52 \pm 005$ than those reported by either HIPASS or 
ALFALFA, though this result is currently based only on 370 galaxies.

\begin{table}[!h]
\caption{Comparison of Major Blind HI Surveys}
\smallskip
\begin{center}
\begin{tabular}{cccccccc}
\hline
\noalign{\smallskip}
Survey & Beam      & Area   & $\delta_v$    & rms$^a$ & V$_{med}$ & N$_{det}$ & Ref\\
       & \arcmin & \sqd & \kms & mJy  & \kms    &           &    \\
\noalign{\smallskip}
\hline
\noalign{\smallskip}
AHISS   & 3.3 &    13 & 16 & 0.7 & 4800 &   65 &  $^b$ \\
ADBS    & 3.3 &   430 & 34 & 3.3 & 3300 &  265 &  $^c$ \\
HIPASS  & 15. & 30000 & 18 & 13  & 2800 & 5000 &  $^{d,e}$ \\
AGES    & 3.5 &   200 & 11 & 0.7 & 9500 & 2900 &  $^f$\\
ALFALFA & 3.5 &  6920 & 11 & 1.7 & 8200 & 31000 &  $^g$\\
AUDS    & 3.5 &   2.7 & 5  & 0.08 & 22000 & 200 &  $^h$\\
ZOA     & 3.5 &   300 & 5  & 0.7 & 6500 & 1200  &  $^i$\\
\noalign{\smallskip}
\hline
\end{tabular}
\\
{\small
$^a$ mJy per beam at 18 \kms ~resolution;
$^b$ Zwaan et al. (1997);
$^c$ Rosenberg \& Schneider (2000);
$^d$ Meyer et al. (2004);
$^e$ Wong et al. (2006);
$^f$ projection, based on results of 30\% data release: www.naic.edu/~ages/public;
$^g$ Giovanelli et al. (2005) and projection for high signal-to-noise objects only, based on results of 70\% catalog);
$^h$ based on detection of 102 sources in half of the projected survey coverage (Hoppmann et al. 2015);
$^i$ based on detection of 61 sources over 15 sq. deg. section out of 300 sq. deg. projected coverage (McIntyre et al. (2015).
}
\end{center}
\label{tab:blindtab}
\end{table}

After the pilot surveys AHISS and ADBS, HIPASS was completed with 
great success. It covers 3/4 of the whole sky, detecting about 5000 sources at a median
redshift $cz\simeq 2800$ \kms, for a sky density of one detection every 6 deg$^2$.
Two major surveys have been completed with the ALFA feed array at Arecibo: ALFALFA
and AGES. The statistics of AGES is difficult to compare with that of other surveys,
since that survey is a collection of selected fields of specific interest, such as galaxy cluster
fields (e.g. A1367) and the near fields of objects with tidal remnants (e.g. NGC 7332/7339), 
each field being mapped blindly, but characterized by different environmental properties.
The total area mapped by AGES is about 200 deg$^2$, for a detection rate of $\sim$15 
sources per deg$^2$, partly resulting from the deep sensitivity limit of 0.7 mJy and partly 
to the selection of {\it a priori} known to be significantly overdense regions. ALFALFA 
covers a solid angle of $\simeq$7000 deg$^2$ with a sensitivity of 1.7 mJy at the 18 \kms
~spectral resolution. The current ALFALFA catalog, which covers 70\%
of the survey area, contains 27,000 sources, thus the complete catalog will exceed
the count of 30,000 sources, for a detection rate of about 4.5 sources per deg$^2$, an
increase in the HIPASS detection rate per unit area by a factor of $\sim 20$.
With a source median redshift of $cz \simeq 8000$ \kms, ALFALFA is the first blind HI
survey approaching coverage of a cosmologically fair volume, for sources at its median
redshift. ZOA is a survey aimed
to inspect connectivity across the galactic plane, of large-scale features in the 
galaxy distribution. With a nominal sensitivity about 2.5 times better than ALFALFA's, it should
yield a significantly larger detection rate per deg$^2$ and median velocity. However its
detection rate is 4.2 deg$^{-2}$ and its median $cz$ is less deep than ALFALFA's. This
is in part due to a less efficient bandpass calibration technique, forced by running
''piggyback'' on a pulsar survey. AUDS is the deepest of the HI surveys to date. It
maps two roughly antipodal sky regions for a total solid angle of 2.7 deg$^2$. The
first data release reports the detection of 102 galaxies over 1110 hours of telescope
time, with a sensitivity level $\sigma_{rms}$ of 80 $\mu$Jy and a median 
$cz \simeq 22,000$ \kms. The low efficiency of the detection rate in this case is
due to the bad conditions of RFI) at
frequencies lower than $\sim$ 1350 MHz ($z > 0.05$), as we discuss in 
section \ref{rfi}. 

The low efficiencies of ZOA and AUDS illustrate the impact of circumstances
not considered by the scaling relations drawn in this section. On the positive side,
it emphasizes the synergies in allowing different surveys sharing simultaneous
use of the telescope, as in  the cases of ZOA riding piggyback on a pulsar survey
and the interstellar HI survey TOGS doing so with ALFALFA. On the negative side the
observing mode imposed by the pulsar survey on ZOA may not be optimal for the latter.
In the case of AUDS, the noisy spectral environment of Arecibo underscores the
importance of site selection for the new generation facilities in Western Australia,
China and South Africa.

Other blind surveys of
somewhat lesser impact are discussed in Henning et al. (2000), Minchin et al. (2004),
Lang et al. (2003), Davies et al. (2004). Prominent targeted surveys with relatively
small samples but high parametric content, carried out with HI synthesis arrays, are
THINGS (Walter et al. 2008), LittleTHINGS (Hunter et al. 2012), FIGGs (Begum et al.
2008) and SHIELD (Cannon et al. 2011).

\begin{figure}
  \includegraphics[width=1.0\textwidth]{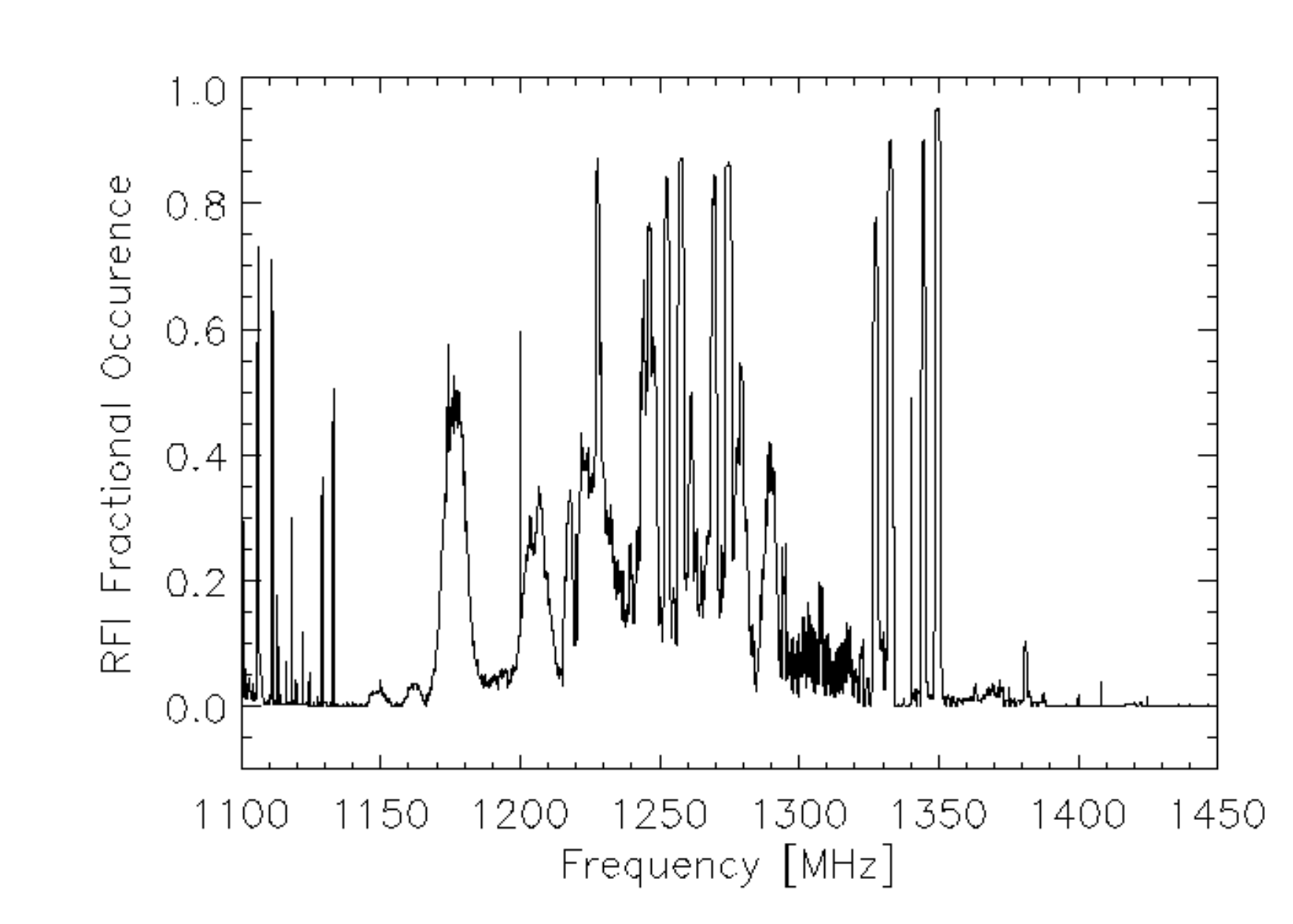}
\caption{\footnotesize {L-Band RFI at the Arecibo Observatory. 
The fractional occurrence of RFI, i.e. the fraction of the time
that observations are severely affected by RFI at any given frequency
at the Arecibo Observatory, sampled with a resolution of 0.1 MHz. The
21cm line rest frequency is 1420.4 MHz. Credit: P. Perillat (2015).
}}
\label{fig:rfi}    

\end{figure} 

\subsection{RFI}
\label{rfi}
The worst enemy of a spectro-photometric, extragalactic HI survey is radio
frequency interference (RFI), such as airport radar, orbit-to-ground data download,
radio stations, cell phone transponders, etc. Alleviating techniques fall short
of clearing the impact of more than a small fraction of RFI sources, so that the
quality of survey data suffers selectively at different epochs and frequency
bands. An example of this impact is shown in Figure \ref{fig:rfi}, which displays
the fraction of the time for RFI occurrence, plotted as a function of frequency
and averaged over the year preceding August 1st, 2015 at the Arecibo Observatory 
(Perillat 2015). The radio L-band spectrum is relatively clean for extragalactic
HI observations at frequencies higher than $\sim 1350$ MHz ($z_{HI}< 0.06$), the
major disturbance being the Global Positioning System (GPS), emitting near 1381 MHz
about 10\% of the time. At increasing $z$ for rhe HI line, the next important RFI 
source is the San Juan airport radar, transmitting between 1325 and 1350 MHz. Below 
1325 MHz a multitude of RFI sources occupy the spectrum. In Fig. \ref{fig:Spaen},
the gap in the source density of souces near $D_{Mpc}=230$ corresponds to
the frequency range obliterated  by the transmissions of the San Juan airport 
radar, located about 100 km from the Arecibo telescope. Apart from GPS, the spectral 
band between 1350 MHz and 1420 MHZ is relatively clean at Arecibo, still allowing 
largely unpolluted observing conditions. However, at 1350 MHz and below  RFI is
very severe.

\subsection{Stacking}
\label{stack}

\begin{figure}
  \includegraphics[width=0.80\textwidth]{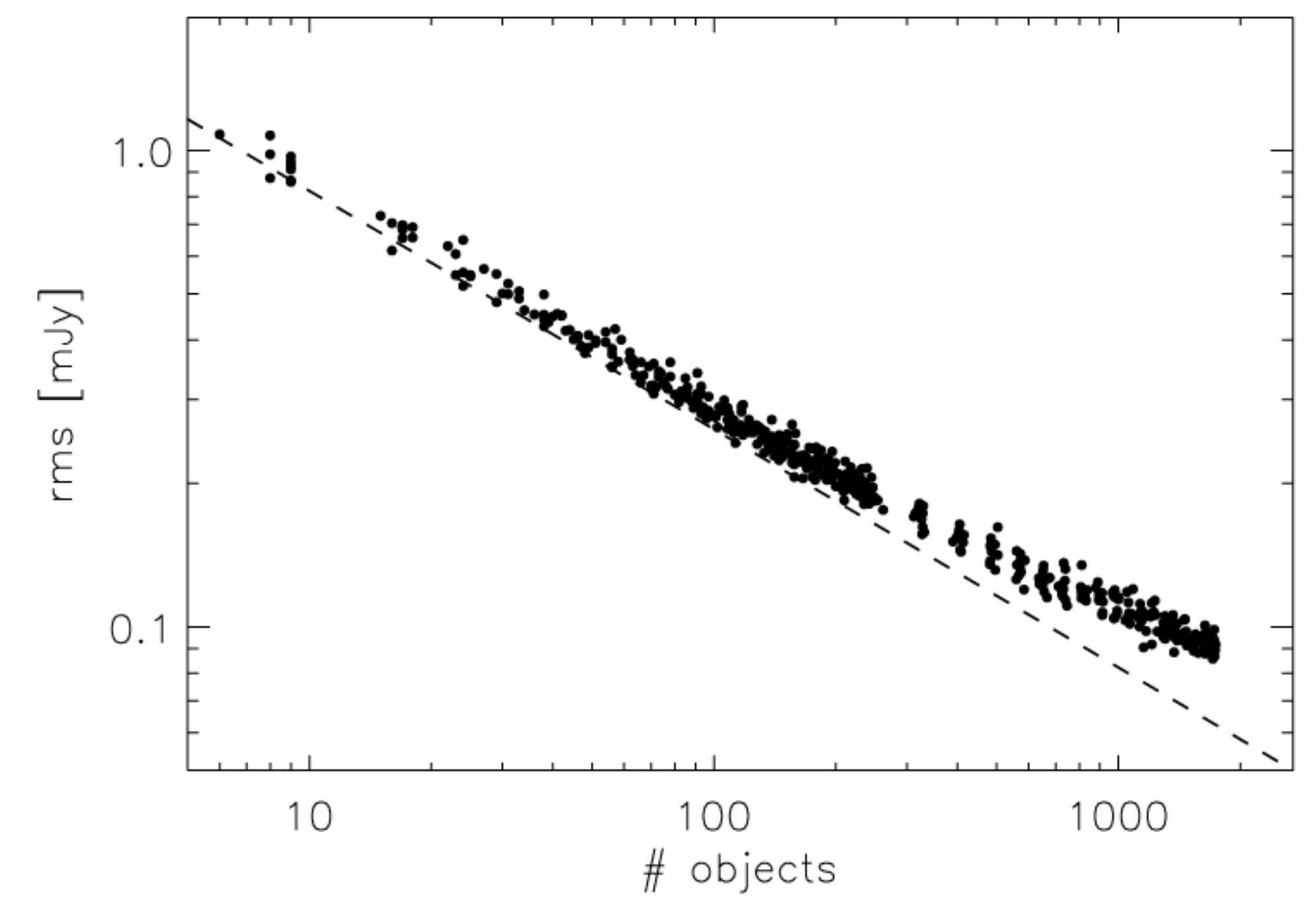}
\caption{\footnotesize {Noise figure of stacked spectra, as a function
of the number of objects co-added. The dashed line is the expected
N$^{-1/2}$ dependence for clear non-Gaussian noise, after Fabello 
et al. (2011).
}}
\label{fig:stack1}    
\end{figure}

An  option available with blind HI survey data is that of obtaining
estimates of average properties of specific categories of sources, to noise
levels significantly lower than those achievable for single sources. Suppose
that a population of sources --- for example early type galaxies --- falls 
within the volume sampled by a survey, but the survey sensitivity is not 
sufficient to individually detect more than a small subsample of the lot. 
If the positions and recession velocities of the sources are known, co-adding 
the survey spectra along the line of sight of a number N --- which can be 
large --- of individually undetected sources can reduce the noise and deliver 
a significantly deeper sensitivity measure of the statistics of the population. 
Before co-adding, each spectrum needs to be shifted in frequency by 
an amount which aligns its recession velocity with that of the other spectra
to be stacked. For the ALFALFA survey, the rms noise of the co-added spectrum 
can be seen to decrease nearly as N$^{-1/2}$, as
shown in Fig. \ref{fig:stack1} ~(Fabello et al. 2011).

\begin{figure}
  \includegraphics[width=1.00\textwidth]{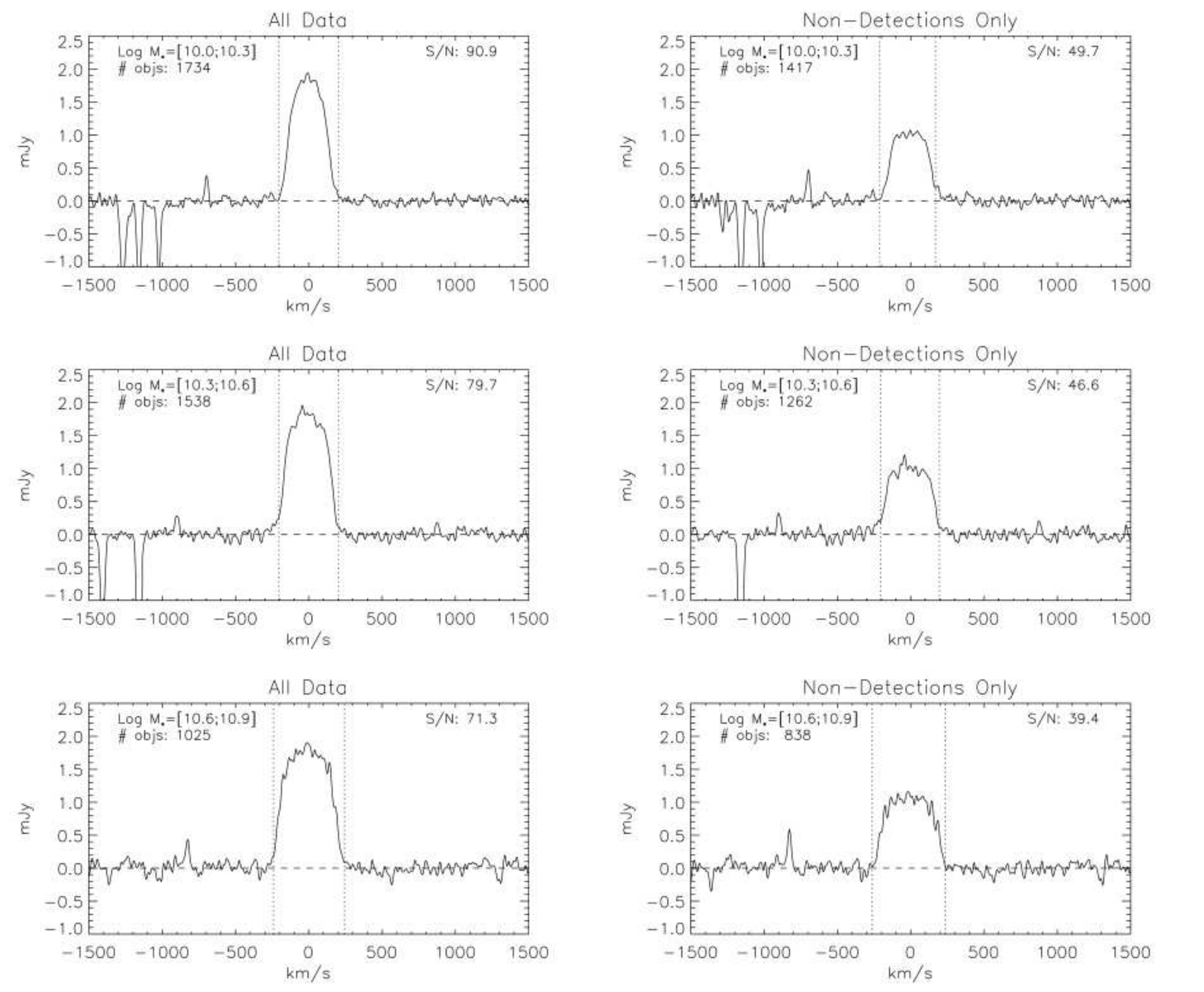}
\caption{\footnotesize {Co-added spectra of an optically selected sample
of early type galaxies are shown, for three stellar mass bins. ALFALFA
detects only about 1/5 of the objects. For each of the mass bins, panels 
to the left show the co-added line profile of all galaxies within that
bin, including those individually detected; panels to the right show the 
line profile obtained by co-adding only the spectra of non-detections.
The inset labels within each box are (i) the stellar mass range over which
spectra have been stacked; (ii) the number of co-added spectra and (iii)
the Signal-to-noise ratio of the co-added spectrum. The narrow features near
$cz\simeq -1000$ \kms ~and below are due to rfi. Credit: Fabello et al. (2011).
}}
\label{fig:stack2}       
\end{figure}

Figure \ref{fig:stack2} illustrates the power of the stacking technique, as
applied to the study of HI content in early type galaxies (ETG) by Fabello et al. (2011).
An optically selected parent sample of 4748 ETGs was extracted from the Sloan
Digital Sky Survey, a subset of
only $\simeq 20$\% of which had been individually detected by ALFALFA. The parent 
sample was subdivided into 5 bins of stellar mass (only the top 3 are shown in the 
figure), and the ALFALFA spectra at the location of each galaxy in the mass bin were 
co-added. Panels in the left column show the result of the co-addition of all galaxies
within the bin, including galaxies individually detected by ALFALFA; panels in
the column to the right show the result from co-adding only spectra with no individual
evidence of emission. E.g. for the top mass bin, which includes 1734 galaxies with stellar
mass between $10^{10.0}$ and $10^{10.3}$ \msun, 1417 show no evidence of individual
emission. However, the co-addition of the 1417 objects shows a clear detection with
a signal-to-noise  ratio of $\sim 50$. As indicated in Section 2.1, ALFALFA can detect
an HI mass of $\sim 10^7$ \msun ~at a distance of 20 Mpc. Figure \ref{fig:stack1} shows that
the technique can reduce $\sigma_{rms}$ by a factor of $\sim 20$; the HI mass detection
limit scales linearly with $\sigma_{rms}$, so ALFALFA can set a limit of $\sim 5\times 10^5$
\msun ~for stacked sources located at 20 Mpc.

The impact of confusion on stacking is discussed, eg. by Delhaize et al. (2013), 
Duffy et al. (2008) and Jones et al. (2015b). The signal fraction derived from
the accumulated emission of targeted, central sources is mixed with a confusion
signal accumulated along the line of sight. The latter is referred to as the ``confused
mass''. Fabello et al (2011) find that a factor of 10-20 below the detection
limit is the most that could be gained by stacking, before non-Gaussian moise becomes 
dominant, as shown by Fig. \ref{fig:stack1}. The confused mass grows rapidly with
redshift of the survey, dominating stacked spectra for $z>0.1$ in surveys carried
out with single dishes.

\section{Properties of the HI Source Population}
\label{sec:properties}

The fundamental properties of individual galaxies correlate strongly:
massive galaxies are brighter and larger than less massive ones,
but they also tend to be redder, of higher metallicity and dustier.
At its simplest, the Hubble sequence illustrates
how strongly galaxy morphology correlates with metrics associated with color which
in turn reflect the history of star formation.
The huge digital dataset produced by the Sloan Digital Sky Survey has
enabled the robust determination of uniformly measured properties 
to place on a quantitative basis the interrelationships among stellar mass, 
optical luminosity, mass-to-light ratio, star formation rate, mean stellar age, 
size, metallicity, velocity dispersion and rotational velocity for the
$z=0$ Universe.  These observational relations for
the local galaxies serve as the end-point for models of the evolution of galaxies over
cosmic time, giving rise to the ``galaxy main
sequence'', i.e. the tight correlation at any given redshift between a galaxy's
star formation rate and its stellar mass. Most of these studies focus on 
massive galaxies with log M$_{star} > 10.5$ the majority of which lie on the
``red-sequence'' and are no longer forming stars. 

In contrast to the wide area optical/IR surveys, the HI surveys sample
a very different population of galaxies, largely excluding the red and
dead galaxies. Not all galaxies contain (detectable) HI and it is important 
to understand the resultant bias on the HI source population. After reviewing 
the clustering properties of the HI source population and its distribution
as a function of HI mass, in this section we discuss our fundamental understanding
of the HI-selected population with emphasis on how it strongly differs
from an optically selected one.

\subsection{The HI Auto-correlation Function}
\label{sec:autocorr}

The morphology-density relation which describes the clustering
properties of galaxies is more often statistically quantified 
by the measurement of correlation functions. In its simplest
form, the autocorrelation function in real space $\xi(r)$ counts
pairs of objects as a function of their separation vector, which
is derived from a component $r_p$ in the plane of the ky and one 
along the line of sight $\pi$. Redshift space distortions acting
on different scales need correcting in order to yield a reliable
$\xi(r)$. A power law function is fitted, $\xi(r)=(r/r_\circ)^{-\gamma}$.
Guzzo et al. (1997)
estimated the power law parameters as a function of morphological type:
$r_\circ = 8.35\pm 0.75 h^{-1}$ Mpc and $\gamma= 2.05\pm 0.09$ for
ellipticals and $r_\circ = 5.55\pm 0.45 h^{-1}$ Mpc and 
$\gamma= 1.73\pm 0.08$ for Sb and earlier spirals; for later spirals
and irregulars, they obtained $r_\circ = 4.05\pm 0.70 h^{-1}$ Mpc and 
$\gamma= 1.5\pm 0.12$. Defining the relative bias parameter $b$ between
elliptical and early spiral populations as
$\xi_E(r)=b^2 \xi_S(r)$, 
they get $b=2.0\pm 0.4$ with only a very mild dependence on scale $r$.
The spiral population is antibiased with respect to the elliptical one. 

The optical counterparts of HI-rich galaxies are very frequently blue, low optical
surface brightness systems, which are known to favor low density environments. Thus,
it is not surprising that HI-selected samples exhibit similar clustering properties.
With important SKA precursor experiments being built with the goal of characterizing
the HI source populations at intermediate redshifts, it is important to establish
soundly the statistical properties of those objects at $z=0$. The HIPASS and
ALFALFA surveys have recently been used for the derivation of the autocorrelation
function of the HI source population. Meyer et al. (2007) and Basilakos et al. (2007)
both agreed in identifying the HIPASS HI sample as the extragalactic population with 
the weakest clustering known; 
the two groups found respectively best 
fits of $3.5\pm0.3$ and $3.3\pm0.4$ $h^{-1}$ Mpc for $r_o$
and $1.47\pm0.08$ and $1.4\pm0.2$ for $\gamma$. While Basilakos
et al. (2007) found that, within the HI population, high mass sources clustered more strongly,
Meyer et al. (2007) found no statistical difference between high and low mass subsamples.
Using the ALFALFA $\alpha.40$ catalog,
Martin et al. (2012) found $r_o=3.3\pm0.3$ $h^{-1}$ Mpc and $\gamma=1.51\pm0.09$, a very good match with the
results based on the HIPASS data. 
The HI selected population is thus antibiased with respect to all optical
samples.

\begin{figure}
  \includegraphics[width=0.50\textwidth]{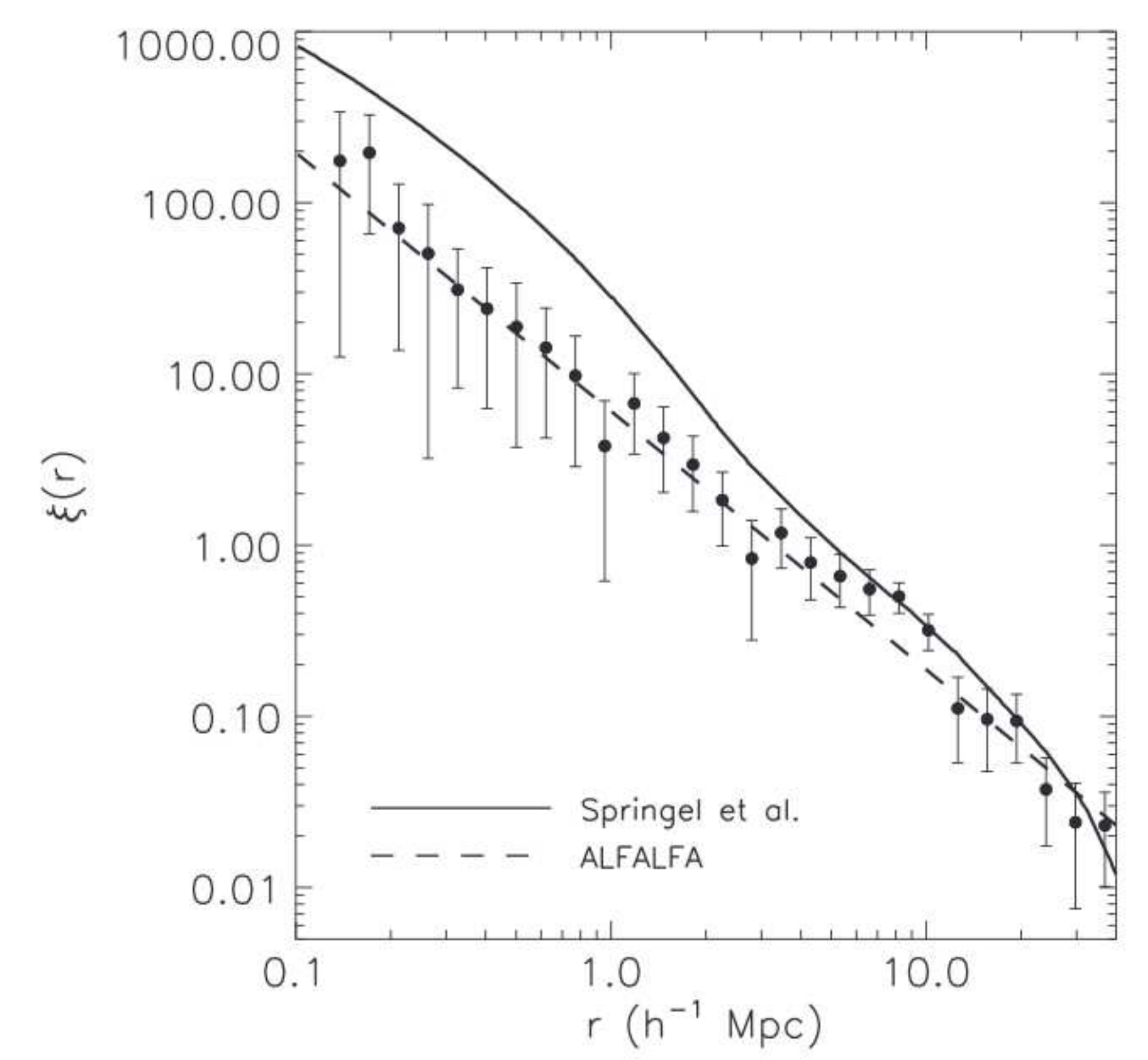}
  \includegraphics[width=0.50\textwidth]{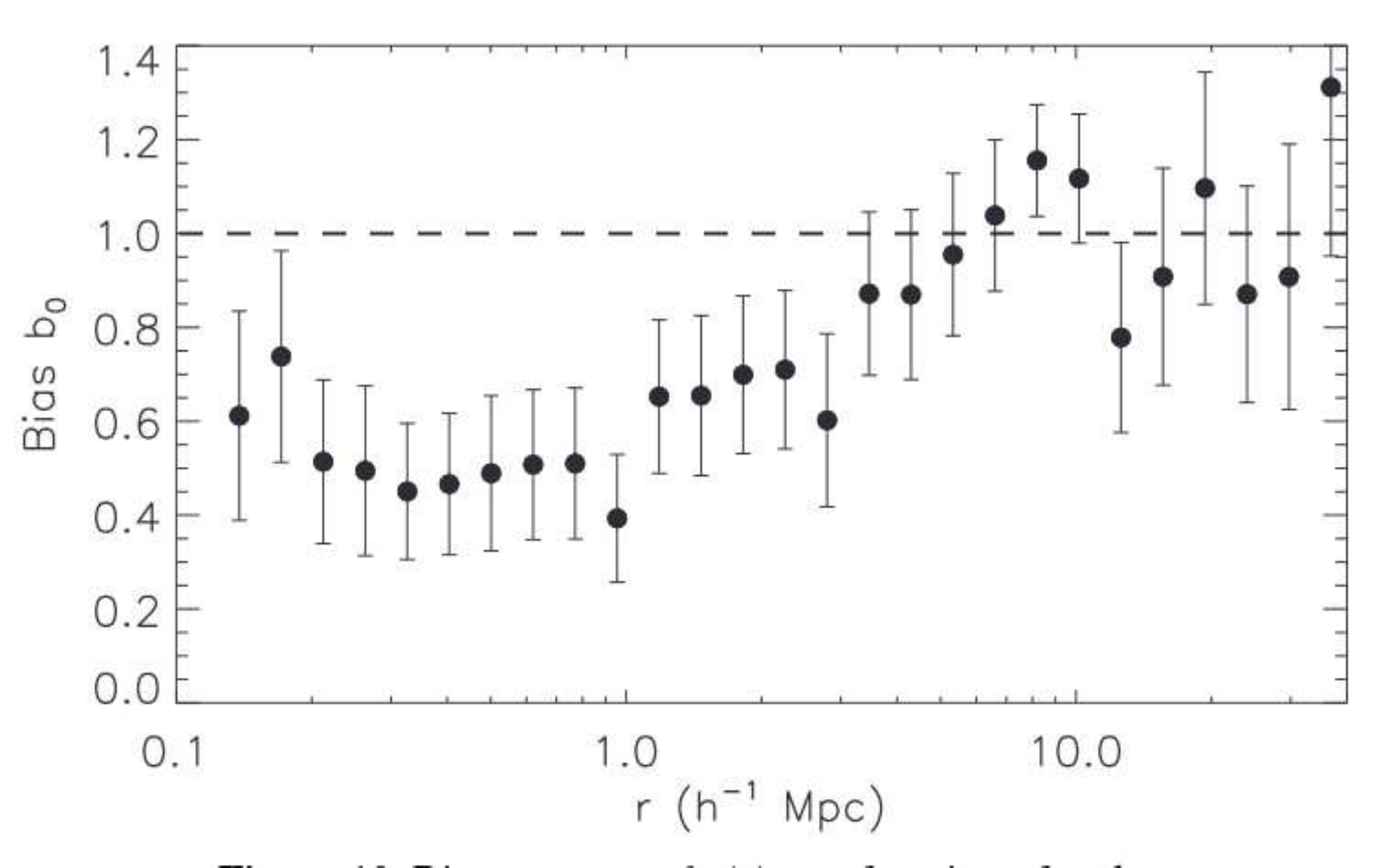}
\caption{\footnotesize {{\bf Left} Real space autocorrelation function for the dark
matter halos from the Millennium simulation (solid line) and for the HI $\alpha.40$
ALFALFA sample (symbols with error bars).
{\bf Right}: Bias parameter  plotted as a function of scale. Both Figures from 
Martin et al. (2012)
}}
\label{fig:xi}       
\end{figure}

It is convenient to compare the autocorrelation function
of the $\alpha.40$ catalog with that of the dark matter halos as produced by \LCDM
~simulations. Then, the bias parameter is shown by Martin et al.(2012) as
a function of scale in the left hand panel of figure \ref{fig:xi}.
The HI galaxy population is antibiased with respect to that of the 
dark matter. The left hand panel of figure \ref{fig:xi}
shows the autocorrelation of the ALFALFA $\alpha.40$ catalog and its best fit model
(dashed line). The solid line is the real-space autocorrelation function of the dark
matter halos $\xi_{DM}(r)$. The right hand panel of figure \ref{fig:xi} shows the
bias parameter as a function of scale. Dark matter is clearly more clustered than the
HI source population, but the bias is scale dependent. HI selected galaxy samples appear
to be severely antibiased on small scales, but only weakly antibiased on large scales.

\subsection{The HI Mass Function}
\label{sec:HIMF}

The distribution of low mass dark matter halos in \LCDM ~simulations can be 
fit by a Schechter function of slope  $\alpha =-1.8$. However, the best fit
slope to the faint end of the optical luminosity (or the stellar mass) function 
of galaxies exhibits a shallower behavior, with slopes typically between -1.0 and 
-1.3. Likewise, the early determinations
of fits to the HI mass function (HIMF) returned low mass slopes which appeared
shallower than that of dark matter halos, both in the case of optically selected
samples (Springob et al. 2005) and in that of samples derived from blind surveys 
(Rosenberg \& Schneider 2000). Using the HIPASS source catalog of 4315 HI sources, 
Zwaan et al. (2003, 2005) obtained $\alpha=-1.37$. However, the HIPASS catalog
contained  only 44 objects less massive than $10^8$ \msun, making the 
determination of the slope uncertain. The ALFALFA catalog $\alpha.40$ contained
10452 high signal-to-noise detections, of which 329 are less massive than $10^8$
\msun. Moreover, with a median redshift of $cz \simeq 8000$ \kms, vs. the $\simeq 2800$
of HIPASS, ALFALFA samples a cosmologically fair volume which is significantly deeper 
than that of HIPASS. The ALFALFA HIMF had a best-fit Schechter function of 
$\alpha=-1.33\pm 0.02$, very close to the value of Zwaan et al. (2005), a 
characteristic ``knee'' mass $M^*=10^{9.96\pm 0.02}$ \msun ~and a normalization parameter 
$\phi^*=4.8\pm0.03\times 10^{-3}$ Mpc$ ^{-3}$ dex$^{-1}$. The early mismatch between
\LCDM ~and actual galaxy counts is now reliably confirmed: there is no detected
population of galaxies making the baryonic (stellar plus gas) mass function approach
the abundance of low mass dark mater halos. 

\begin{figure}

  \includegraphics[width=0.50\textwidth]{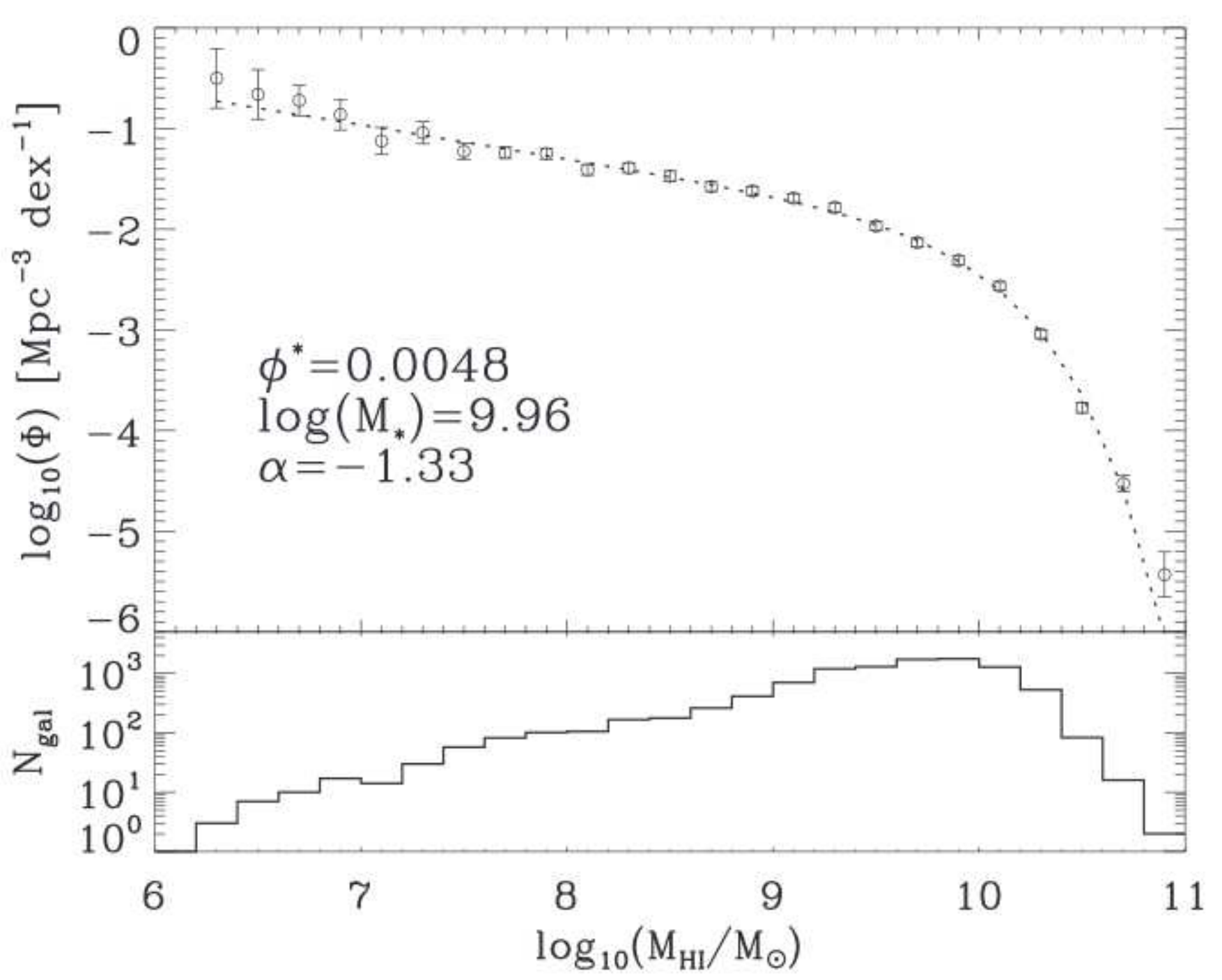}
  \includegraphics[width=0.50\textwidth]{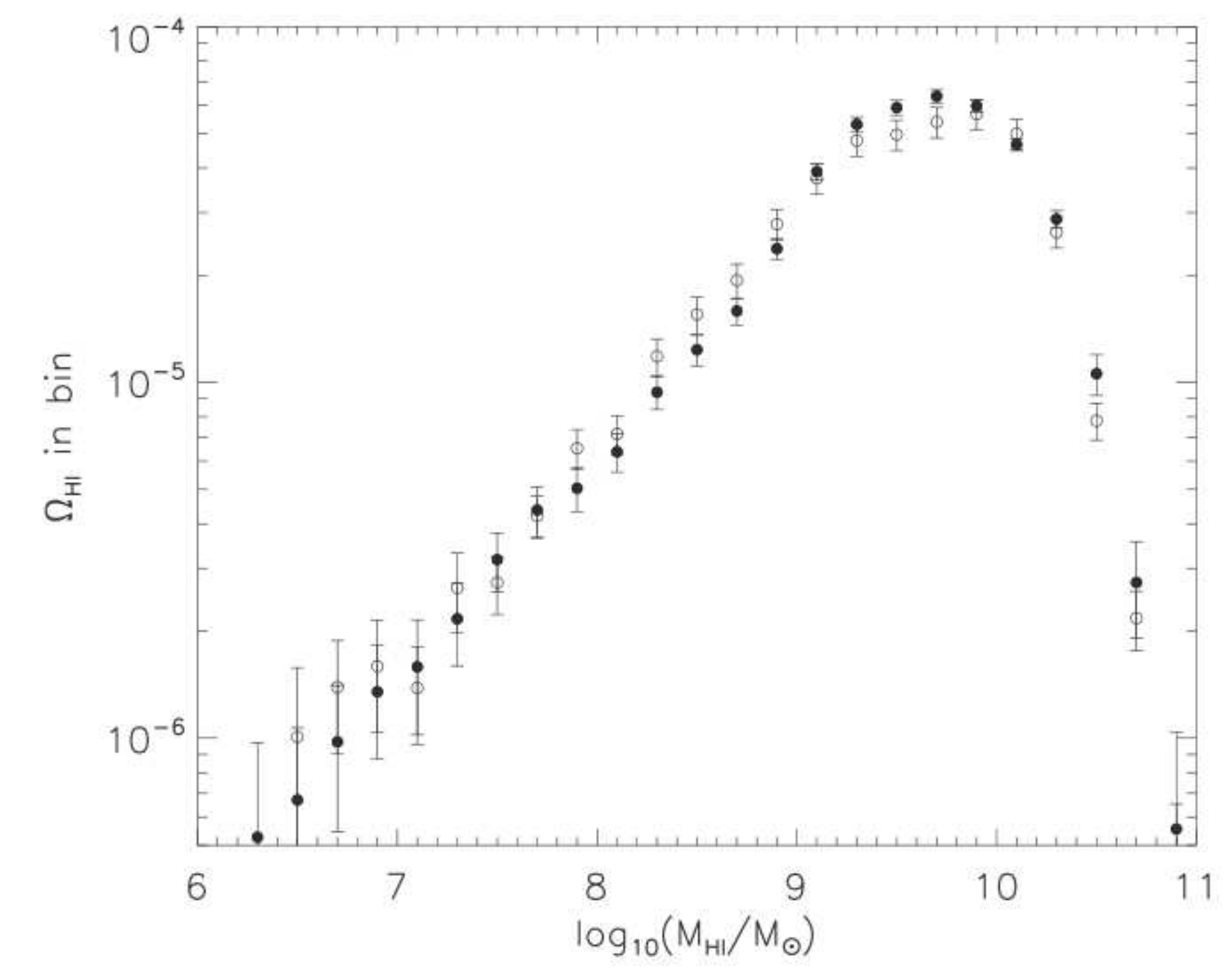}

\caption{\footnotesize {{\bf Upper left:} The ALFALFA HIMF, based on the $\alpha$.40 catalog,
as obtained via the SWML method, after Martin et al. (2012). Inset are the best fit parameters 
of a Schechter model. The lower plot is a histogram of the number of HI sources, per mass bin,
on a logarithmic scale.
{\bf Right hand:} Histogram of the contribution to $\Omega_{HI}$ of galaxies binned by HI mass.
Filled symbols refer to calculations of the HIMF via the $1/V_{max}$ method, while unfilled ones
refer to using the SWML method. The overall density of HI in the $z=0$ Universe is mainly
contributed by galaxies with $9.0 < \log (M_{HI}/M_\odot) <10.0$. Figure after Martin et al. (2012).
}}
\label{fig:HIMF}       
\end{figure}

The HIPASS and $\alpha.40$ HIMFs however
disagree in the counts of sources with the highest masses: at $\log M_{HI}/M_\odot=10.75$
ALFALFA finds $5\times$ more objects than HIPASS and at $M_{HI}\simeq 10^{11}$ \msun ~the gap
widens to one order of magnitude. The $\alpha.40$ HIMF obtained via the stepwise maximum 
likelihood (SWML, Efstathiou et al. 1988) method is
shown in figure \ref{fig:HIMF}. Li et al. (2012) have derived a photometric estimator of the 
HI mass fraction $M_{HI}/M_*$ using the ALFALFA catalog and applied it to a larger sample
of $~24000$ SDSS galaxies in the redshift range $0.025<z<0.05$ and found that the bias 
parameter is a function of both scale, as found by Martin et al. (2012, cf. Figure \ref{fig:xi}), and HI mass
fraction, galaxies with higher mass fraction being significantly more antibiased.

The resulting cosmic density of HI in the $z=0$ Universe is, according to 
ALFALFA, a fraction $\Omega_{HI}= (4.3\pm 0.3)\times 10^{-4}$ of the critical density. While the
HI-rich dwarf galaxy population dominates the cosmic density by number, the total
density of HI at $z=0$ is largely contributed by galaxies with $9.0 < \log (M_{HI}/M_\odot) < 10.0$,
as shown in the right hand panel of Figure \ref{fig:HIMF}. In a paper presenting the result of
half of their survey (AUDS) data, Hoppmann et al. (2015) report the detection of 102 galaxies with
redshifts up to $z=0.16$ (median $z=0.065$). This sample yields an HI mass function statistically 
compatible with those obtained by HIPASS and ALFALFA. By splitting the data into subsets of high and 
low redshift, they suggest a possible decrease in $|Omega_{HI}$ towards the upper end of the redshift
range. Combining their results with those of other surveys, they estimate 
$\Omega_{HI}=(2.63\pm0.10)\times 10^{-4}$.

\begin{figure}
  \includegraphics[scale=.65]{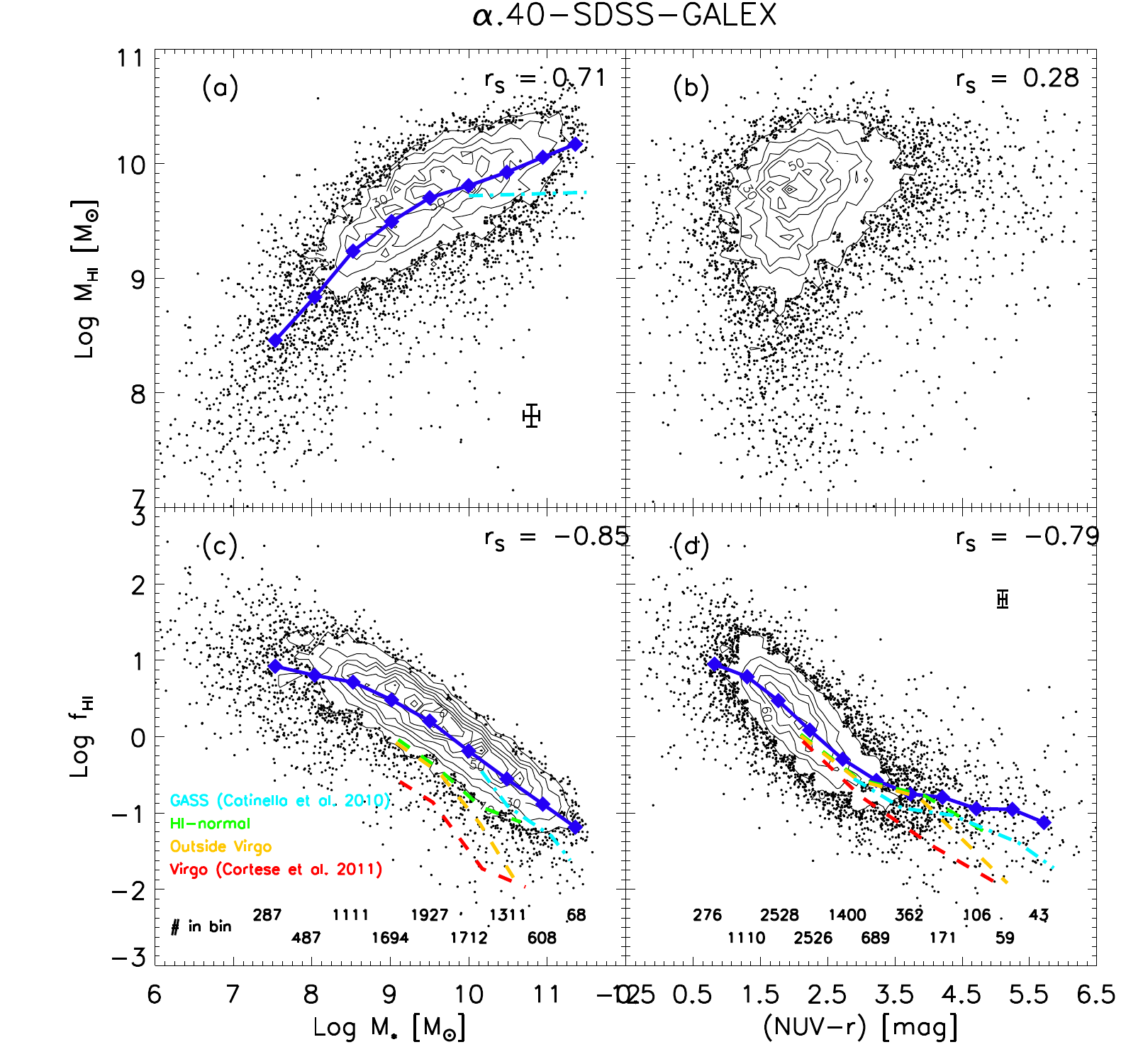}

\caption{Scaling relations for the HI mass and atomic gas fraction $f_{HI}=M_{HI}/M_*$
as functions of stellar mass $M_*$ and (NUV-r) color. Contours trace the distribution
of galaxies commonly detected by ALFALFA-SDSS-GALEX, with the lowest contour
at a level of 20 galaxies per grid cell, contour steps of 20 counts per cell; 
at lower densities, individual
points are plotted. Lines in several panels trace mean relations from
the GASS survey Catinella et al. (2010) and the local survey of Cortese et al. (2011).
From Huang et al. (2012b).
}
\label{fig:HIscale}       
\end{figure}

\subsection{What galaxies emit HI?}
\label{sec:bias}
In combination with the large, deep, homogeneous SDSS database, the major HI
surveys allow determination of how the HI content and gas fraction vary with
comparable properties that describe the stellar component. Figure \ref{fig:HIscale},
from Huang et al. (2012b), shows the scaling relations between the HI mass and
HI fraction (M$_{HI}$/M$_*$) with the stellar mass and (NUV-r) color for the
$\alpha.40$ catalog. Contours denote the densely populated
regions (cells with 20 or more galaxies), below which individual points are plotted. 

Huang et al. (2012) find that a transition in the star formation (SF) properties 
appears to occur around stellar masses of M$_* \sim 10^{9.5}$ M$_{\odot}$.
Below that break point, the slope of the star-forming sequence changes
and the dispersion in the specific star formation rate (SSFR) increases; in fact, at low masses, the
SF appears to be strongly regulated by the atomic gas, and the star formation
history is bursty and non-continuous. Over the same
volume, the
population detected by HI surveys has higher overall SFRs and SSFRs at
fixed stellar mass. The evidence suggest that the HI-rich galaxies are
less evolved than an optically-selected population, either because
there is some bottleneck in the conversion of atomic to molecular gas
or that the overall star formation law in HI-dominated galaxies is 
different and relatively inefficent. Curiously, Huang et al. (2012)
find that, at a fixed stellar mass, galaxies with higher gas fractions
appear also to be hosted by dark matter halos with higher than average
angular momentum (as measured by the observational equivalent of the
spin parameter $\lambda$.

\subsection{What galaxies don't HI surveys see?}

Canonical understanding states that early-type galaxies (ETGs) contain very little
cool gas and that disk galaxies in the cores of rich clusters are highly HI-deficient
relative to comparable galaxies in the field, notably as a result of ram
pressure stripping. 

Early type galaxies (ETGs) in general are ``red and dead'', assumed having long ago
converted their atomic gas supply into stars. Of course, that stereotype
is a gross oversimplification with many ETGs, particularly those outside
clusters showing modest amounts of HI Grossi et al. (2009). In fact,
Serra et al. (2012) summarize the results of HI synthesis mapping, principally
with the WSRT, of 166 nearby ETGs also contained in the ATLAS$^{3D}$ survey.
The detection rate of this survey is $\sim$10\% for ETGs which are Virgo members and 
$\sim$40\% for ETGs outside Virgo. Serra et al. (2014) point out that the HI morphologies
of ETGs, as well as their masses, are surprisingly diverse, with about 25\% of
the field ETGs hosting quite massive (up to 10$^9$ \msun, extended HI disks or rings.
Those authors suggest that current simulations do not predict the observed HI
properties of ETGs.

\subsection{HI-deficient galaxies}
\label{sec:HIdef}
That spiral galaxies in clusters are much poorer in their gas content
than field dwellers of comparable morphological type and size, and that
their gas disks are truncated due to ram pressure stripping, has been known 
for more than 40 years (Haynes et al. 1983; Giovanelli \& Haynes 1983;
Boselli \& Gavazzi 2006 and 
refs. therein). More recent studies have investigated the problem in
greater detail (Solanes et al. 2002; Vollmer et al. 2004; Chung et al.
2009), bringing more clarity to the understanding of the physical process. 
The study of HI deficiency is
however only marginally aided by blind, shallow surveys like ALFALFA 
and HIPASS: deep, targeted HI observations are required in order to
reliably measure HI deficiency. The one aspect that large, blind surveys
can contribute effectively to the measurement of HI deficiency is in the 
determination of the standard of ``HI normalcy'', by measuring the properties 
of large samples of galaxies found in various stages of isolation (Toribio et al. 2011).
In Virgo, the sensitivity of current HI instruments
is adequate to probe the lower mass population.
Boselli \& Gavazzi (2014) review the full body of evidence concerning
the formation and evolution of low mass red galaxies in clusters, especially 
Virgo thanks to its vicinity, confirming the general conclusion
that ram pressure stripping of infalling dwarfs is the dominant 
process, leading to the quenching of star formation in clusters. They also show
that sequential harassment will bear an impact depending on where
galaxies have resided in a cluster throughout their history. Based on ALFALFA,
Hallenbeck et al. (2012) show that some HI-bearing Virgo dEs
likely have accreted their gas only recently.

\begin{figure}
\includegraphics[scale=.75]{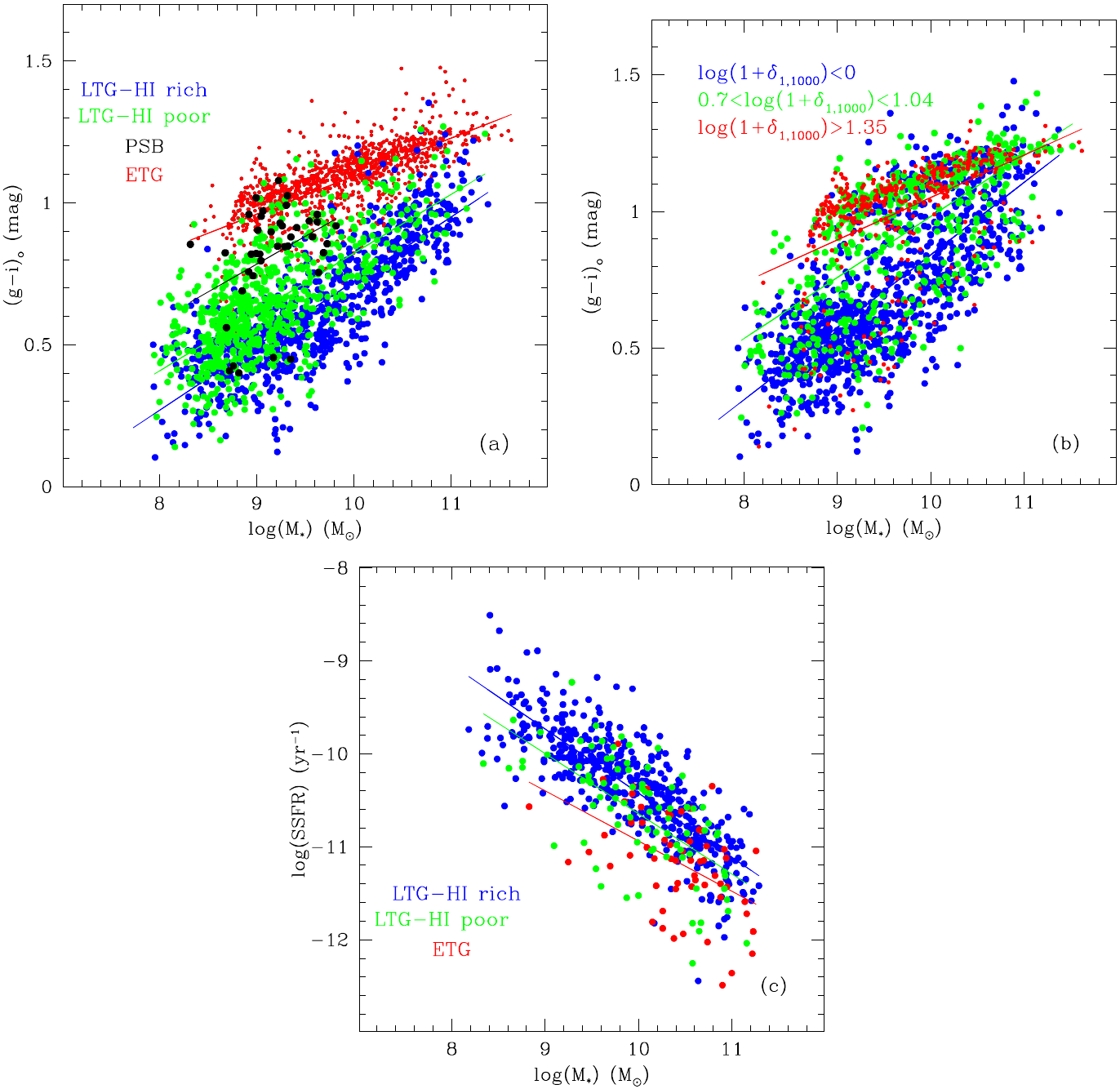}
\caption{Four-category sequence in the HI gas-color relationship for 
galaxies selected on the basis of morphology, gas content and nuclear
spectral properties: late type galaxies (LTG) which are HI-rich or
HI-poor, post-starbursts (PSB) and early-type galaxies (ETG).
From Gavazzi et al. (2013).
}
\label{fig:gav13}
\end{figure}

A number of recent studies aimed to identify the
threshold in environmental density when gas stripping begins to significantly impact 
the gas content. Making use of a rich collection of multi-wavelength
data from SDSS and H$\alpha^3$, a narrow band imaging survey
of a flux-limited HI sample extracted from ALFALFA, 
Gavazzi et al (2013) postulate the identification of an evolutionary
sequence in galaxies manifesting itself towards progressively earlier 
morphology, decreasing gas content and decreasing star formation
activity, in the neighborhood of rich clusters. In the dwarf regime 
($M_* < 10^{9.5}$ \msun), they identify a four step sequence, from
(i) HI-rich late type galaxies, (ii) HI-poor late type galaxies,
still exhibiting nuclear star formation, to (iii) HI poor galaxies without
either extended or nuclear star formation but with post starburst signature,
and (iv) redder early type galaxies showing no gas or star formation on all
scales, as shown in Fig. \ref{fig:gav13}. They propose ram pressure as the 
most likely mechanism in producing this transformation. For Coma, they 
estimate the infall in the cluster at a rate of approximately
100 galaxies with $M_* > 10^9$ \msun ~per year.

\subsection{The HI view of star formation laws}
\label{sec:SFL}

Star forming galaxies show a tight correlation between their stellar masses
and the star formation rate (SFR), known as the ``star formation main sequence''.
This relation appears to hold for massive galaxies (log M$_* >$ 10)
over at least the redshift range $z <$ 3 (Rodighiero et al. 2011; Genzel et al. 2015).
In local star-forming galaxies containing various amounts of HI, SF occurs 
in massive, dusty, dense and cold
molecular clouds. So we are faced with the question: Where does the atomic 
gas fit in? In particular, it is not clear how star formation on sub-kpc scales,
strongly correlated with the molecular gas surface density, can
lead to the global scaling relations such as those seen in Figure
\ref{fig:gav13}.

Kennicutt \& Evans (2012) present a thorough review of current understanding of the
process of star formation on the scale of galaxies and how star formation
couples with the physical characteristics of the interstellar gas. The
accumulating body of observational data on the stellar, dust and multi-phase
(cold, cool, warm, hot) gas on sub-kpc scales enables tests of the most
commonly-used relation between the surface density of star formation 
and of the gas, the so-called Kennicutt-Schmidt Law $\Sigma_{SFR} \propto
\Sigma_{gas}^N$, with $N\sim 1.4$. In combination with the accumulating
observational datasets acquired at other wavelengths, HI synthesis maps 
enable the exploration of the behavior of the spatially-resolved and 
dust-corrected star formation rate with the changing mix of atomic and 
molecular gas.

As discussed in detail also by Kennicutt (2010), 
Schruba et al. (2011) have combined the HERACLES IRAM-30m/CO
study with the VLA/HI THINGS to show quantitatively that the atomic and molecular 
phases relate to star formation in very different ways. 
Different SF laws appear to apply in different regimes
of $\Sigma_{gas}$: the HI-dominated medium below log $\Sigma_{gas} \sim 1$ M$_{\odot}$ pc$^{-2}$;
the H$_2$-dominated at 1 $<$ log $\Sigma_{gas} < $ 2.3 M$_{\odot}$ pc$^{-2}$,
and the starburst regime at higher gas densities: c.f. Fig12 of Kennicutt \& Evans (2012).
Overall, the HI surface density
appears to have little to no impact on the local $\Sigma_{SFR}$;
even where the HI dominates,
the SFR correlates more closely with the molecular gas. 
It appears that gas is consumed more rapidly if the molecular fraction
is high. The normal process of conversion from atomic to molecular
appears to stall in the HI-dominated regions of galaxies, thereby reducing the SF
efficiency even in the presence of significant amount of gas.

To first order, the HI-dominance is also reflected in the metallicity of
the gas. Low metallicities are seen in dwarf galaxies and the outer 
disks of late-type spirals. Only in rare instances does CO detection take
place in very low metallicity regimes (Elmegreen et al ~2013).
Since dust grains play critical roles both in 
hosting the surface formation of molecules and in shielding them from 
ambient UV photons, it is not surprising that such HI-dominated
regions are characterized both by low gas surface-densities 
($\Sigma_{gas} < 1$ M$_{\odot}$ pc$^{-2}$) and higher gas-to-dust ratios.
There the global efficiency of star formation
$\Sigma_{SFR}/\Sigma_{gas}$ is relatively low and uncorrelated with
$\Sigma_{gas}$ (Bigiel 2008; Kennicutt \& Evans 2012). 
Simulations have shown that H$_2$ may form even in gas with few
or no heavy elements (e.g. Krumholz 2013 and references therein),
enabling a path to star formation even in the most unprocessed environments.
THINGS and HALOGAS (Heald et al. 2011) have assembled a rich dataset of
very deep HI imaging on a growing number of large nearby systems with significant
column density sensitivity, to allow searches for in-falling clouds
and streams of material. To date, no study has identified accretion
at a level sufficient to provide continuous fueling of the current
star formation rate. A caveat about these works however reminds us that
column density limits can only be quoted for beam-averaged quantities,
that is, under the assumption of a smooth distribution of gas 
which fills the radio telescope
beam and which refers only to the neutral gas. Clumpiness or high fractional
ionization can significantly skew the interpretation, or even worse, hide
the true gas content. Much work remains to be done, especially at levels
of truly low column density and on the smallest scales.

With the advent of the Atacama Large Millimeter Array, studies are just beginning
to probe the properties of the bulk of the star-forming galaxies at high redshift.
First suggestive studies, notably PHIBSS (Tacconi et al 2013), have begun to explore
the redshift range $z \sim$ 1-2 using interferometers to identify 
massive, turbulent gas disks at that epoch. Without the sensitivity of the
future SKA, HI studies at this point lag far behind. As a pilot program
the HIGH$z$ survey of Catinella \& Cortese (2015) have reported direct detections of 39
actively star forming disks at Arecibo, selected from the SDSS, at $z >$ 0.16, in 
observing conditions more RFI-benign than current ones. 
Interestingly, the HIGH$z$ galaxies are unusually blue and gas-rich, and lie
in relatively isolated fields. Catinella \& Cortese (2015) argue that they are indistinguishable
from the high gas fraction, high HI mass HIghMass galaxies identified in the
ALFALFA survey (Huang et al. 2014; Hallenbeck et al. 2014), but are very different from the PHIBSS molecular-dominated
turbulent disks. The HI and H$_2$ populations cannot be linked unless the PHIBSS galaxies contain
a large HI reservoir or are not representative of the general disk population at the
higher redshifts. It will be a long time before HI facilities have the
capability to detect individual galaxies similar to the PHIBSS disks
at redshifts $>$ 1, but the stacking of sources in deep fields such as proposed for
the MeerKAT LADUMA project (see Section \ref{future}) should provide constrains on the atomic gas content
of the turbulent disks.

\subsection{The Stellar and Baryonic Mass Functions and Baryonic Deficit}
\label{bardef}

The measurements of the stellar mass/luminosity function (SMF) go back 
decades (Felten 1977), but only recently wide field, multi-band 
spectro-photometric optical and near infrared surveys have made 
it possible to determine the SMF from samples of many tenths of
thousands of galaxies (e.g. Baldry et al. 2012 
and refs. therein), extending its estimate over the mass range 
between $10^7$ and $10^{12}$ \msun. The low mass end of the SMF 
is fit by a power law of slope -1.3, in contrast with the steeper power
law of index -1.8 predicted by \LCDM ~for the halo mass function (HMF).

\begin{figure}
  \includegraphics[width=0.90\textwidth]{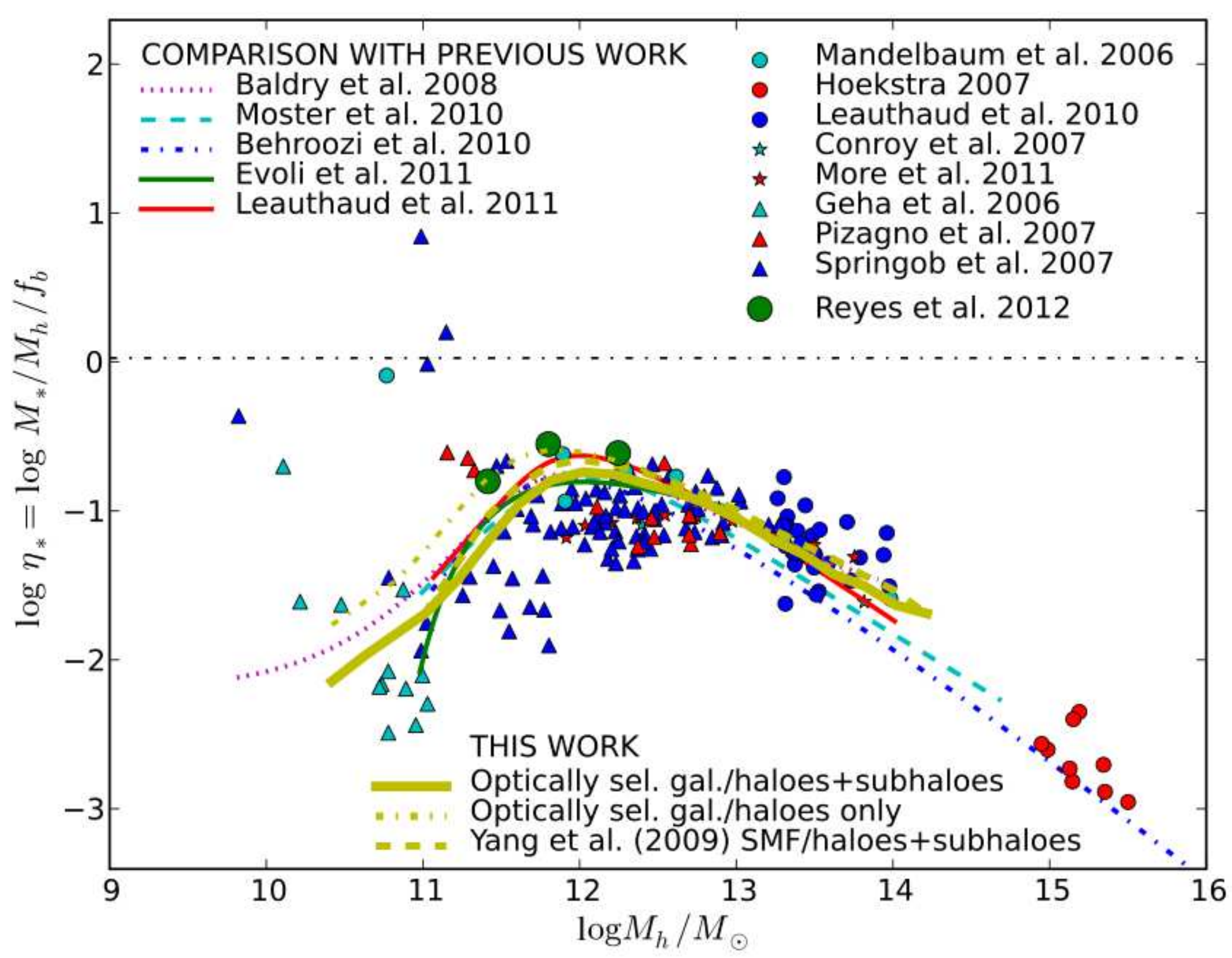}

\caption{\footnotesize { 
Logarithm of the ratio of stellar mass to halo mass, plotted as a function of the 
logarithm of halo mass, normalized by the cosmic baryon fraction. The thick yellow line
results from applying the AM technique to the SMF and the HMF. Other dashed or dotted lines
correspond to the same function as reported, using AM, in the references listed top left
of the figure. The plotting symbols, as listed top right of the figure, identify
measurements of the stellar to halo mass ratio whereby the halo mass was estimated
by a variety of different techniques. Of particular interest are the three filled,
green circle data points, corresponding to measurements of the mean halo mass, via the
signature of weak lensing, in the stacked images of three subsets, of several thousand 
images each  (Reyes et al. 2012). Figure credit: Papastergis et al. (2012).
}}
\label{fig:bardef}      
\end{figure}

The technique of ``abundance matching'' (hereafter ``AM'';
Marinoni \& Hudson 2002; Papastergis et al. 2012, 2015 and refs. therein), 
has been used extensively in the literature, to link the properties 
of an observed galaxy function, e.g. the SMF, with those of the HMF.
By pairing the most massive galaxy with the most massive halo, the second most 
massive galaxy with the second most massive halo, and so on, AM produces 
the average function relating stellar mass with halo mass $M_h(M_*)$, i.e.
the most likely mass $M_h$ of the halo hosting a galaxy of mass $M_*$. Several
different derivations of that relation are shown in Figure \ref{fig:bardef} 
(Papastergis et al. 2012), as solid, dashed or dotted lines. The green filled 
circles show measurements of the stellar to halo mass ratios; the latter 
were obtained via stacked weak lensing observations (Reyes et al. 2012); 
the other symbols are from studies that used kinematic data of individual 
galaxies to estimate halo masses. The y-axis of the plot is $\eta_*=\log [(M_*/M_h)/f_b]$,
i.e. the stellar to halo mass ratio is normalized by the cosmic baryon fraction 
$f_b\simeq 1/6$. Different techniques concur in indicating a relatively low 
efficiency of galaxies in retaining their baryons, or in converting them into 
stars, across the full stellar mass spectrum. The fraction $\eta_*$ of baryonic 
mass which a galaxy converts into stars is significantly 
less than one. It peaks near $\eta_*\sim 0.25$ for a (Milky Way-like) halo mass 
of $\sim 10^{12}$ \msun, and drops rapidly below 0.1 on both 
sides of the peak.

\begin{figure}
  \includegraphics[width=0.70\textwidth]{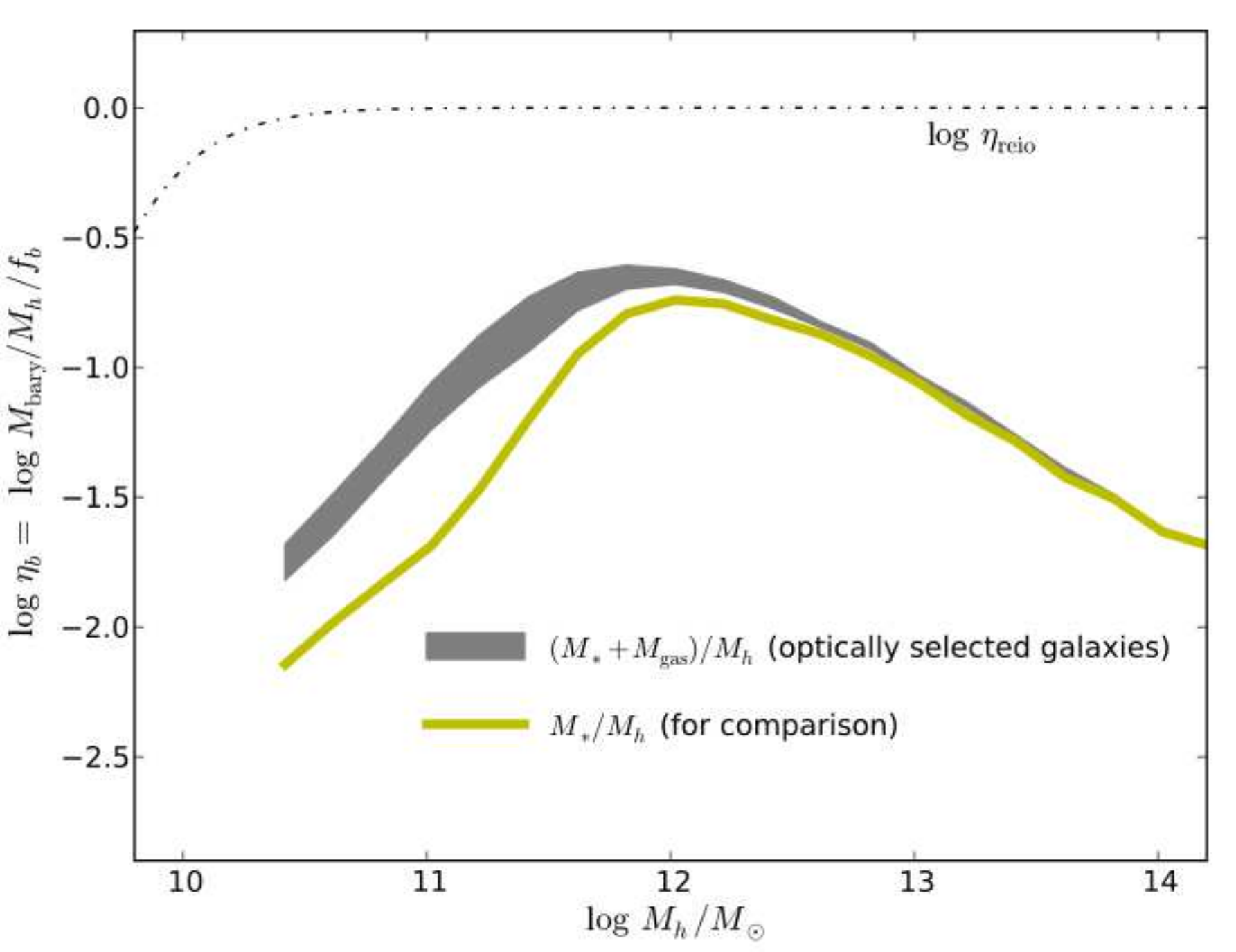}

\caption{\footnotesize {In grey shading, the logarithm of the ratio of baryonic to 
halo mass is plotted as a function of the logarithm of halo mass, normalized by the 
cosmic baryon fraction. For comparison, the yellow line shows the ratio of stellar 
to halo mass, same as in Figure \ref{fig:bardef}. Figure credit: Papastergis et al. (2012).
}}
\label{fig:bardef2}       
\end{figure}

Stellar mass, however, is the main baryonic component only in relatively massive 
systems. The ratio $f_{HI}=M_{HI}/M_*$, where $M_{HI}$ is the HI 
mass, increases with decreasing stellar mass from its peak, becoming the dominant 
baryonic component for systems with $M_*< 10^{10}$ \msun ~(Huang et al. 2012a). 
It is customary to refer to the quantity $M_b=M_* + 1.4\times M_{HI}$ as the ``baryonic 
mass'' of a galaxy, where the factor 1.4 accounts for the Helium mass. We do so, 
with the understanding that: (i) the molecular mass 
is a small fraction of the stellar mass when $M_* > M_{HI}$ and it is a small 
fraction of the atomic gas mass when $M_* < M_{HI}$ (Saintonge et al. 2011),
and (ii) the mass of a hot corona may be the dominant --- but generally not measured ---
baryonic component in
massive systems. Figure \ref{fig:bardef2} shows the correction from $\eta_*$ (the 
thick yellow line), to $\eta_b$, the value inclusive of both stars and atomic gas 
(the grey shaded region, after Papastergis et al. 2012). 

Optically dark and
almost dark HI sources are detected by ALFALFA, alleviating the baryon deficit 
from the magnitude shown by the yellow line in Figure \ref{fig:bardef2} to that
indicated by the grey  one. However,
that alleviation is not sufficient to overcome most of the observed baryonic
deficit, as we discuss further in Section \ref{tbtf}.

Having directly measured HI masses for each individual galaxy
in the derivation of the baryonic mass function is important. 
Obtaining an indirect estimate of the cold gas in galaxies by using
an average scaling  relation between $M_{HI}/M_*$ and $M_*$, 
Baldry et al. (2008) inferred that the increasing gas fraction in 
dwarf galaxies should approximately offset the decreasing stellar-to-halo 
ratio and should result in a roughly constant, bottom-line $\eta_b\simeq 0.1$. 
This spurious result would have implied
that dwarf galaxies are relatively efficient at retaining baryons, but very
inefficient at converting them into stars. As indicated by Kochanek \& White
(2001) and Sheth et al. (2003), this approach is vulnerable; in this case
leading to erroneous inferences on the derived baryonic mass function. The
direct measurement of the baryonic mass made possible by an HI survey
circumvents this vulnerability.

\section{(Almost) Optically-Dark Extragalactic HI sources}
\label{dark}
\vskip -0.15in
An important feature of the ALFALFA data releases is the identification 
of a “most probable” optical counterpart of each HI source, a task 
performed as an integral part of the process of ALFALFA catalog construction 
(c.f. Haynes et al. 2011). Although the vast majority of HI line sources
detected by ALFALFA can be clearly identified with an optical counterpart, 
about 6\% appear to be optically-``dark'' vis-a-vis public, wide field
data bases such as SDSS. A large fraction of those includes the long-known 
high velocity HI clouds (HVCs), a subset of which are the ultra-compact 
high velocity HI clouds (UCHVCs). Excluding HVCs, some of which may be 
extragalactic as we discuss in section  \ref{uchvc}, the extragalactic but 
dark ALFALFA detections amount to about 1\% to 2\% of the full catalog. Sources in this 
small subset fall in three categories: (i) the spectral line does not arise from
an HI transition but rather one of OH megamasers at redshifts near 
$z\sim 0.2$ (Darling \& Giovanelli 2002; Suess et al. 2015), (ii) tidal 
appendages of gas--rich galaxies or remnants of ram pressure events in clusters
and (iii)``something else''.
The source count for the latter category adds up to several dozens. Their 
classification as optically-dark in the ALFALFA catalog is based on examination 
of the SDSS and DSS2 optical images in the vicinity of the ALFALFA HI centroid. 
In some cases, a faint, low-surface brightness stellar counterpart may be visible 
in deeper images than provided by the public surveys; hence the modifier 
``almost-dark'' is adopted. 
We discuss UCHVCs and objects in category (iii) respectively in sections \ref{uchvc}
and \ref{gallery}.

\begin{figure}[t]		
 \vspace*{0.5 cm}
  \includegraphics[width=0.50\textwidth]{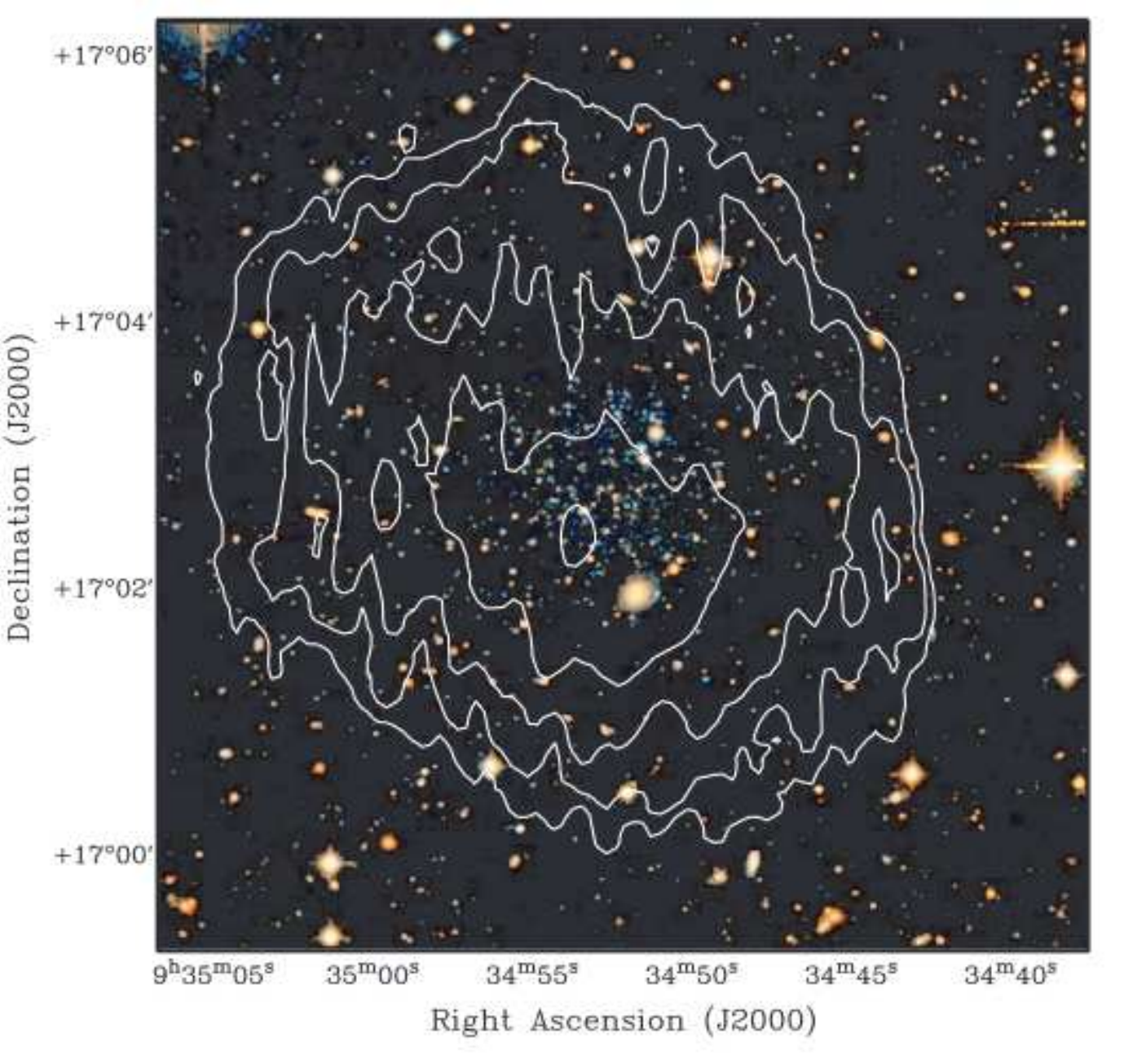}
  \includegraphics[width=0.50\textwidth]{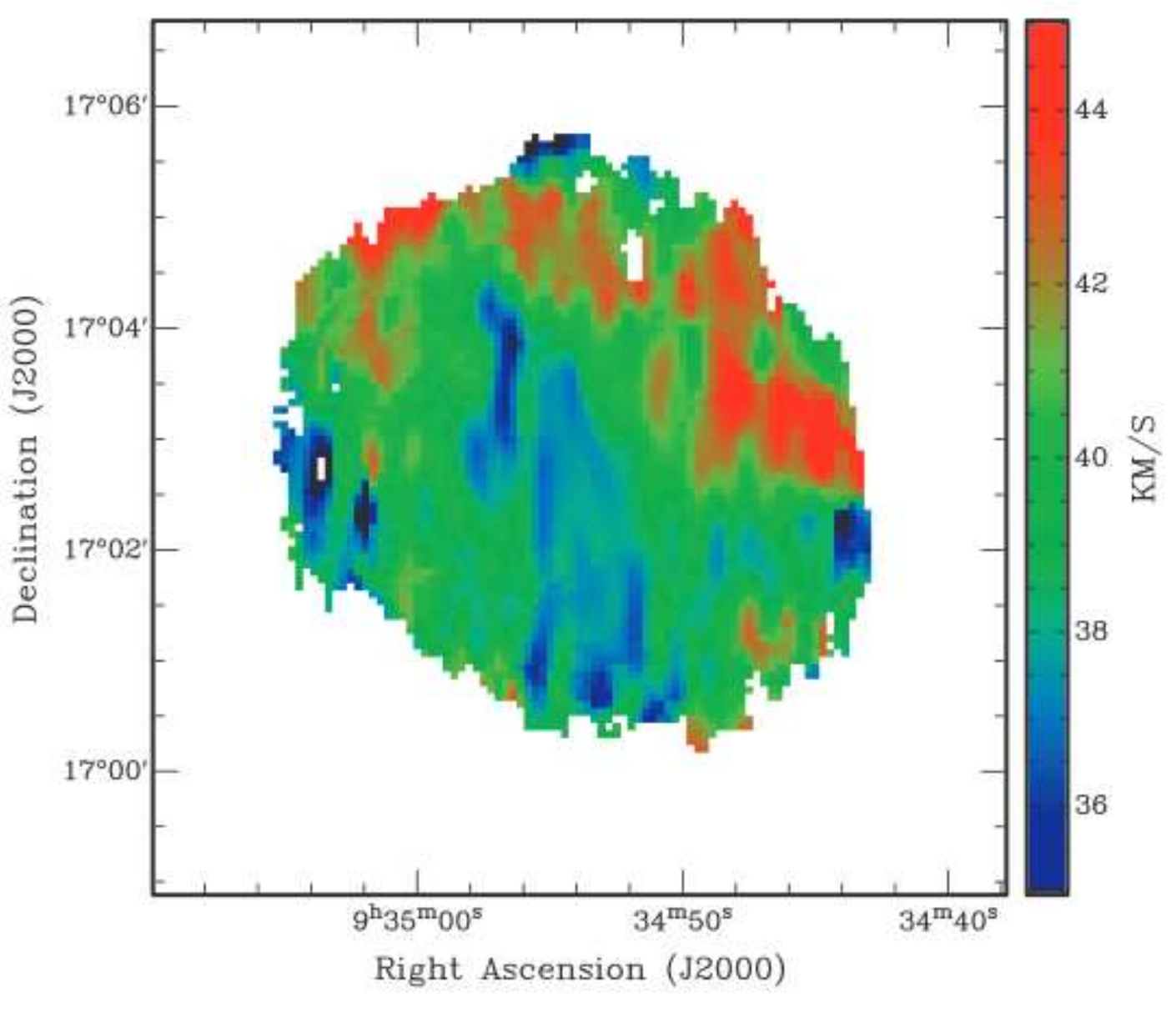}
 \caption{\footnotesize 
	{{\bf Left:} Integrated HI synthesis image of Leo T (HI column density contours at 2.5, 
	10, 20 and 50 $\times 10^{19}$ cm$^{-2}$), superposed on a deep $g-$ and $r-$ image..
	{\bf Right:} Velocity field of Leo T from WSRT data. A hint of rotation? Credit for both
	panels: Ryan-Weber et al. (2008)
}}
   \label{fig:leoT}
\end{figure}

\subsection{Ultra Compact High Velocity Clouds (UCHVCs)}
\label{uchvc}

The idea that compact HVCs may represent the baryonic counterparts of low mass 
dark matter halos, undetected by optical surveys but located within or in the 
near vicinity of the Local Group, was first proposed in two influential papers
by Blitz et al. (1999) and Braun \& Burton (1999). However, the compact HVCs 
identified as possible candidates extracted from early, low resolution surveys 
failed to qualify: if located at Mpc distances, they would strongly violate the
\LCDM ~mass/concentration relation and should have been, but were not detected 
by extant surveys of nearby galaxy groups (Pisano et al. 2008, Sternberg et al. 2002).
The idea was reconsidered more recently (Giovanelli et al. 2010; Adams et al. 2013),
as ALFALFA's angular resolution allowed detection of even more compact HVC sources
than available in the 1990s, labeled ultra-compact HVCs. Convincingly proving
that an ALFALFA UCHVC is indeed the baryonic counterpart of a low mass dark matter 
halo is difficult. It requires higher HI angular resolution maps than provided by 
ALFALFA ($\sim 4'$), in order to identify a possible rotationally supported disk and/or
some indication of distance, e.g. through the detection of an optical counterpart
which would be fainter than the limit of extant optical surveys but bright enough 
to resolve individual stars.

Leo T was found as a stellar overdensity in the SDSS (Irwin et al. 2007). 
Follow-up HI synthesis observations with the GMRT and the Westerbork arrays 
(Ryan-weber et al. 2008) revealed an HI counterpart with a heliocentric velocity
of 35 \kms, also detected by HIPASS and ALFALFA, which would have found it to be
a good UCHVC candidate. At an estimated distance of
420 kpc, its HI mass is $2.8\times 10^5$ \msun, its stellar mass is $0.7\times 10^5$
\msun, its HI radius is 0.3 kpc and the dynamical mass is $>3.3\times 10^6$ \msun.
Its color-magnitude diagram shows evidence of two episodes of star formation:
a currently ongoing one, of approximate age of 200 Myr, as well as an older one,
aged 6--8 Gyr. Figure \ref{fig:leoT} (left panel) shows HI column density contours at the levels
2.5, 10, 20 and 50$\times 10^{19}$ cm${-2}$, superimposed on a $g-$ and $r-$
image, while the right panel shows the velocity field.

AGC 198606, also known as ``friend of Leo T'',  is an ALFALFA-discovered UCHVC located 
16 \kms and $1.2^\circ$ from Leo T. No optical counterpart has yet been found, but 
HI synthesis imaging with WSRT shows signature of a rotating disk with 
$V_{rot}\simeq 14$ \kms ~(Adams et al. 2015). Its distance is not yet known.
Assuming that it is located at 420 kpc,  the same distance of Leo T, it would 
have an HI mass of $6.2\times 10^5$ \msun, an HI radius at the $5\times 10^{18}$
atoms cm$^{-2}$ isophote of 1.4 kpc, a dynamical mass within that radius of 
$1.5\times 10^8$ \msun ~and an upper limit for its $i'$-band absolute magnitude 
of -6.6.

\begin{figure}[t]		
\begin{center}
 \vspace*{0.5 cm}
  \includegraphics[width=0.58\textwidth]{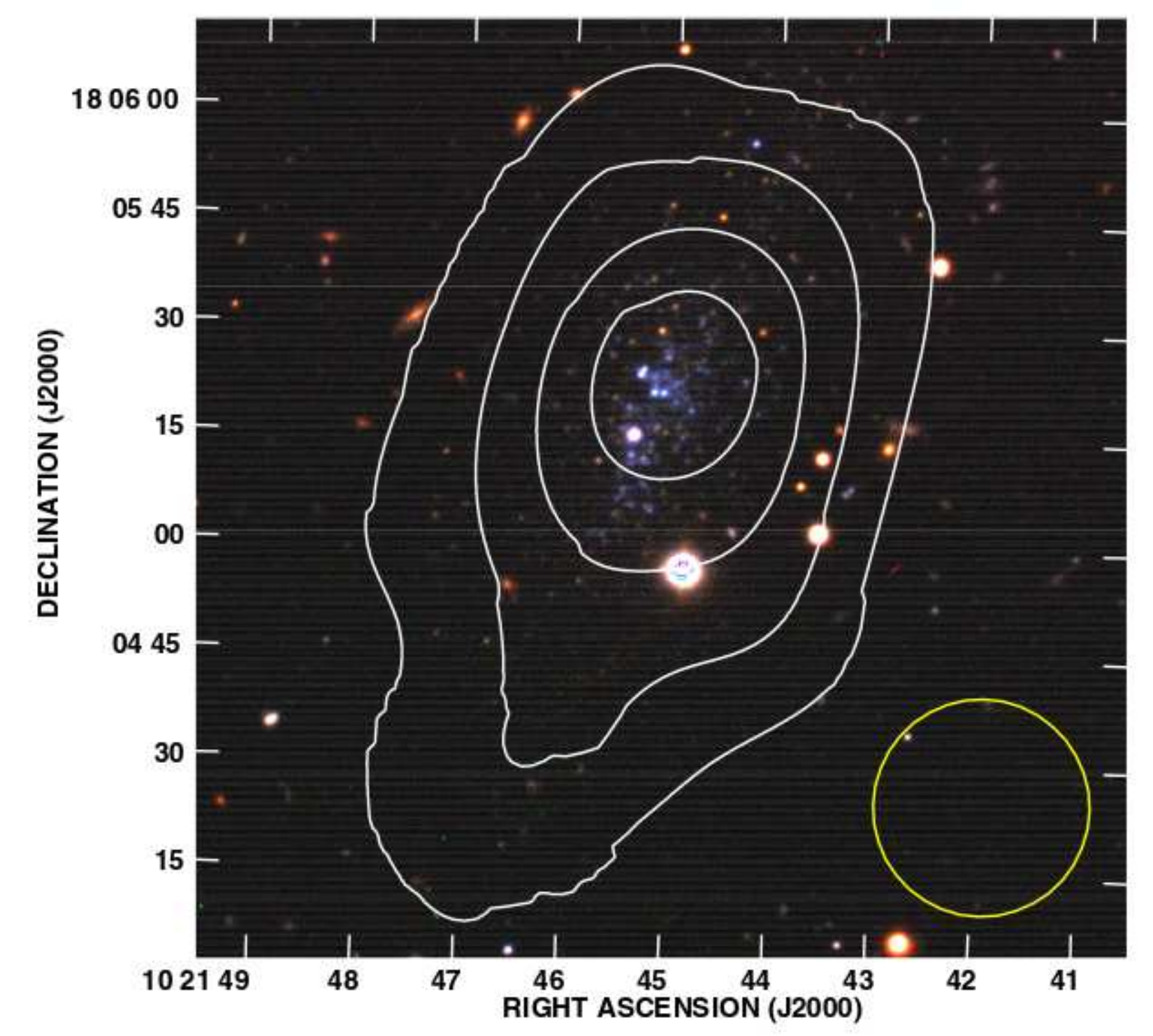}
  \includegraphics[width=0.35\textwidth]{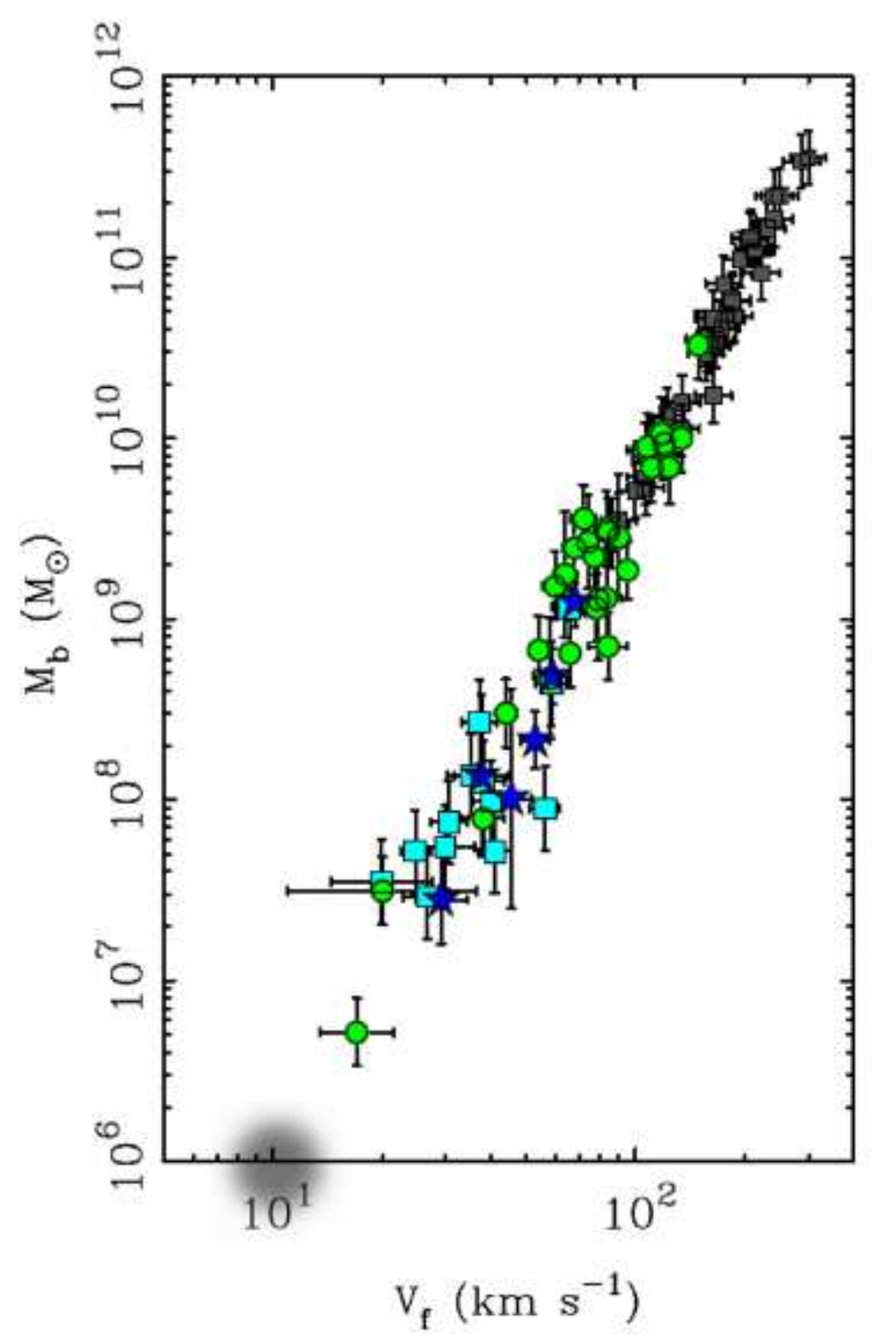}
 \caption{\footnotesize 
	{\bf Left:} VLA-C integrated HI synthesis image of Leo P (HI column
	density contours at 0.5, 1, 1.5 and 2.0 $\times 10^{20}$ cm$^{-2}$)
	smoothed to an angular resolution of 30\arcsec~ (yellow circle), superimposed
	on the WIYN telescope image.
	{\bf Right:} Location of Leo P (fuzzy symbol) on the
	baryonic TF relation. Both figures from Giovanelli et al. (2013), 
	the latter adapted from McGaugh (2012).
}
   \label{fig:LeoP}
\end{center}
\end{figure}

Leo P was discovered by ALFALFA and cataloged as an UCHVC. In the
SDSS it appears as a faint, irregular, lumpy feature, identified as a
compact group of galaxies by McConnachie et al. (2009). However, deep imaging
with the WIYN telescope resolved the optical emission into stars and a bright 
HII region, likely ionized by an O7V or O8V of about 25 \msun ~mass.  
Leo P is located 1.6 Mpc away (McQuinn et al. 2015), has a stellar mass of 
$5.7\times 10^5$ \msun and an HI mass twice as high, an HI radius of half a 
kpc, a dynamical mass of $2.3\times 10^7$ Mpc within that radius and 
($12+\log(O/H)=7.16\pm 0.04$) (Giovanelli et al. 2013, Rhode et al. 2013, 
McQuinn et al. 2013), making it the lowest metallicity, star forming galaxy 
known in the Local Volume (Skillman et al. 2013). With a heliocentric radial 
velocity of 264 \kms, it lies in the near periphery of the Local Group.
The best-fitting star formation history favors ``star formation at the 
earliest epochs, followed by a period of quiescence and a relatively 
constant star formation rate at recent times'' (McQuinn et al. 2015).
HI imaging with the VLA and GMRT revealed a rotating disk with 
$V_{rot}\simeq 10$ \kms ~(Bernstein-Cooper et al. 2014). Figure \ref{fig:LeoP} 
shows VLA-C HI integrated column density contours superimposed
on an optical image taken with the WIYN telescope (left hand panel) and the
location of Leo P in the baryonic TF relation (fuzzy symbol; right hand panel).
This source is a confirmed dwarf galaxy within a minihalo. Had it been $\sim 2.5$
times farther away, it would not have been detected by ALFALFA.

\subsection{Two Examples from the Gallery of Darks}
\label{gallery}

By definition, the baryonic component of systems of category (iii), ``something else''
is dominated by atomic gas, rather than by stars. To add to their exceptionality, some 
of these systems are not ``minihalo'' candidates such as Leo P and Leo T: their HI masses can 
rival that of the Milky Way. However, their HI linewidths tend to be quite narrow, 
typically $< 50$ \kms ~full width at half power, similar to that seen in dwarf galaxies. 
These strange almost-dark ALFALFA sources appear to have been unable to deliver 
a sustained star-formation rate (SFR) at a level that would be compatible with 
their gas content. There is a well-known mismatch between the expected SFR of 
a galaxy, as assessed from the masses of giant molecular clouds gauged to reach
collapse, and its observed SFR. In the case of the Milky Way, the mass
of giant molecular clouds and their estimated free-fall times would
yield a current SFR of $\sim 250$ M$_\odot$ yr$^{-1}$, sustainable
over several Gyr, yet the observed SFR is $\sim$100 times lower
(Zuckerman \& Evans 1974; McKee \& Williams 1997).  Similar results
have been obtained by Wong \& Blitz (2002) for a sample of nearby
galaxies. Theoretical studies indicate that the solution of the
problem might invoke regulation by turbulence, an approach that
appears also to explain scaling relations such as the
Kennicutt--Schmidt law (Krumholz \& McKee 2005 and references therein). 
The objects with extreme $M_{HI}/L_{opt}$ being detected by ALFALFA
exacerbate that mismatch. As a case study, the pair of almost dark
objects named AGC 229384/5 is described next.

\begin{figure}[t]		
\begin{center}
 \vspace*{0.5 cm}
  \includegraphics[width=0.47\textwidth]{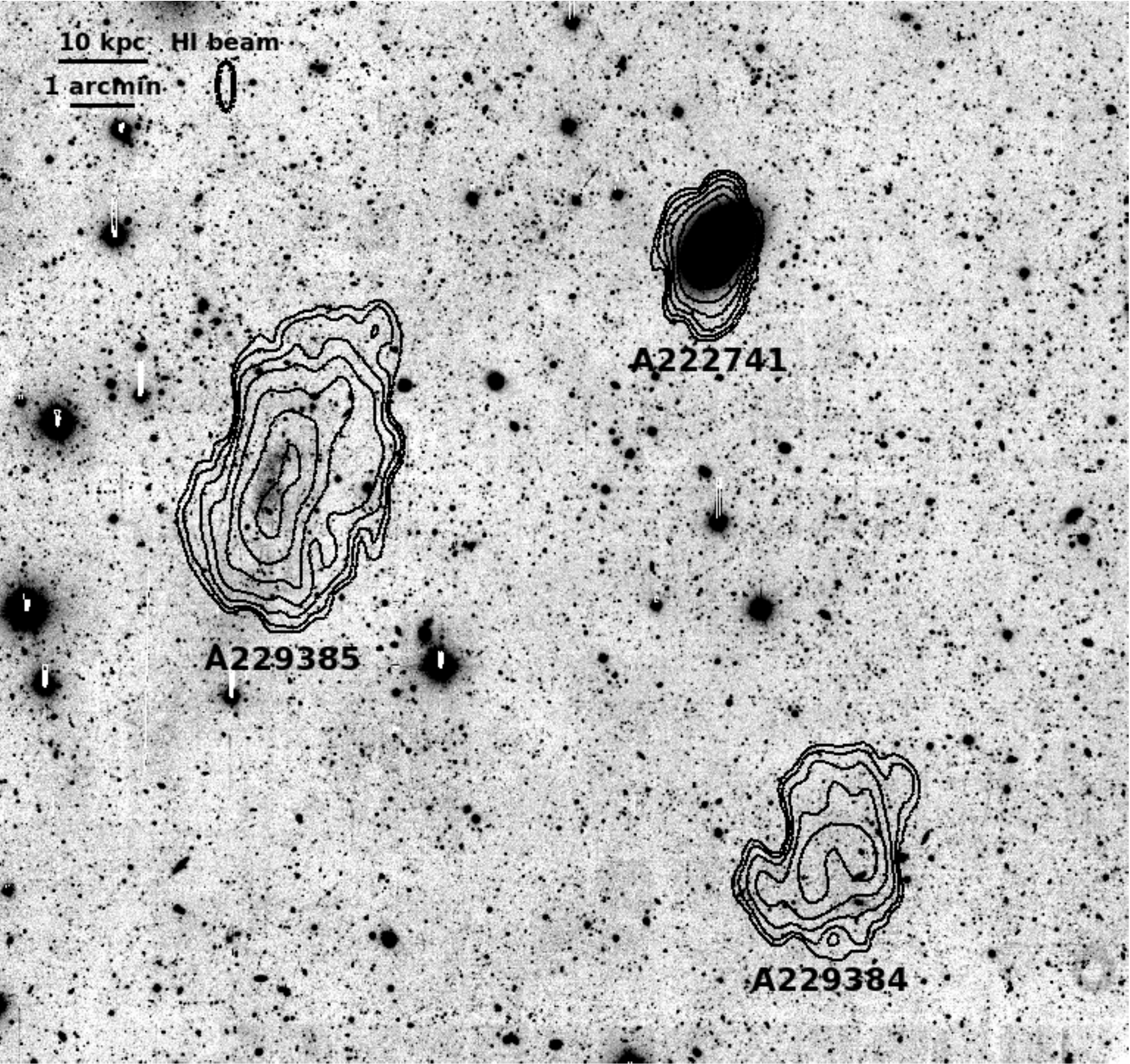}
  \includegraphics[width=0.45\textwidth]{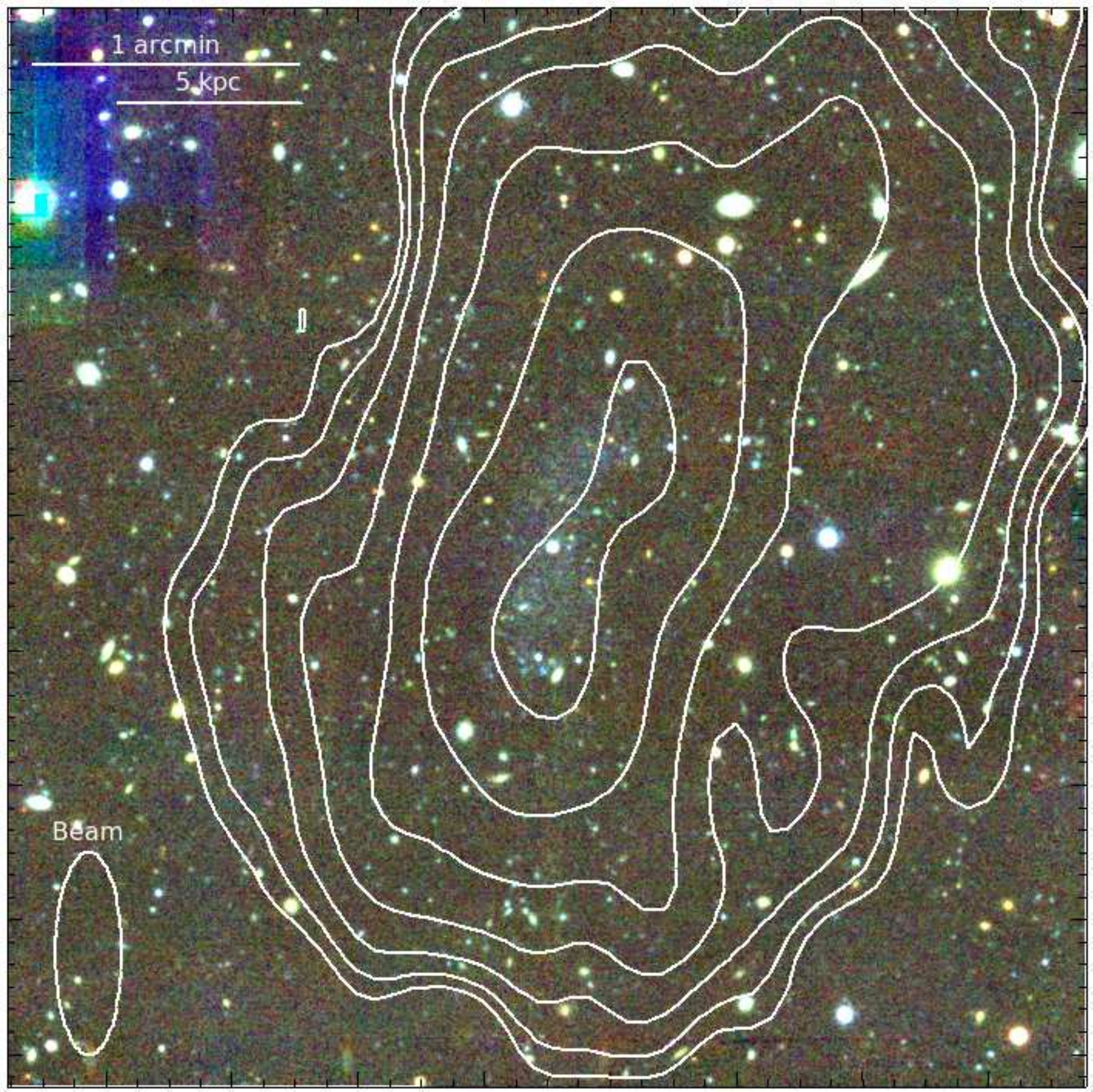}
 \caption{\footnotesize {\bf Left}: A deep $g$, $r$ and $i$ composite image
of a $12$\arcmin~$\times 12$\arcmin~ field containing a pair of
almost-dark ALFALFA sources. Superimposed
are the contours of the HI column density obtained with the WSRT,
at  1, 2, 4, 8, 16, 32, and 64 $\times~10^{19}$~cm$^{-2}$.
Three HI sources are clearly detected. Two of them (labelled AGC~229384 and
AGC~229385) are at $cz$ of respectively 1309 and 1348 \kms;
the third, AGC~222741, is an undisturbed background spiral at $cz=1844$ \kms. {\bf Right:}
A $4$\arcmin~$\times 4$\arcmin~ blow up centered on AGC~229385, showing a faint, blue 
optical counterpart coincident with the HI source. AGC~229384 has no optical counterpart.
Credit: Janowiecki et al. (2015).
}
   \label{fig:comap}
\end{center}
\end{figure}

The left panel of Figure \ref{fig:comap} displays the WSRT HI column density
contours superimposed on an optical image of a field containing several 
ALFALFA sources. The optical image is a composite of three 
45-min exposures in filters ($g, r$ and $i$) made with the
pODI camera at the 3.5m WIYN telescope (Janowiecki et al. 2015). 
The strongest HI source is an undisturbed 
background spiral (CGCG 129-006 = AGC~222741) at c$z$=1884 \kms.
The other two are in the foreground and have no SDSS counterpart.

While AGC~229384, at $cz=1309$ \kms, has no detected optical counterpart even 
in the composite WIYN image, AGC~229385, at $cz=1348$ \kms, has a faint, blue 
stellar counterpart of $g=19.27$ and peak $g$ surface magnitude of 26.6 mag 
arcsec$^{-2}$, evident in the zoomed up right hand panel of Figure \ref{fig:comap}. 
A flow model which accounts for local inhomogeneities of the cosmic density field
yields a most probable distance of $\sim$25 Mpc for the two objects, although 
in this region, projected not far from the virial radius of the Virgo cluster,
the uncertainty on the distance error may be as high as 50\%. The properties of 
both (almost) dark objects are extreme. With $M_{HI}=10^{8.8}$ \msun, only
a few times smaller than that of the Milky Way, AGC~229385 has a $M_{HI}/L_g=44$,
and a stellar mass of only $\sim10^7$ \msun. Moreover, with a linewidth $W_{50}=34$
\kms and an HI radius of 14 kpc, the dynamical mass within that radius, 
$M_{dyn}=10^{9.0}$ \msun, is comparable to that of its HI mass. Were it not for 
its current star-forming activity, AGC~229385 might have escaped optical detection 
altogether. At 25 Mpc, HI mass, radius and linewidth for AGC~229384 would be
respectively $10^{8.3}$ \msun, 7 kpc and 27 \kms; the dynamical mass within the
HI radius would be $10^{8.5}$ \msun. The apparent shallowness of the potential well
of these two sources is exceptional, given their size. Small tidal disturbances
would be able to remove much of their gas, so their current properties may be
made possible by their relative isolation. A suggestive possibility would describe
them as long quiescent, low mass halos, recently rejuvenated by accretion of a 
substantial mass of intergalactic gas, just inducive to starting a star-forming 
episode. As hypothesized by Salpeter \& Hoffman (1995)
in explaining the low redshift Ly$\alpha$ absorbers, these may be ``vanishing
Cheshire cat'' galaxies.

\section{Missing Satellites, etc.: Too Big To Fail?}
\label{tbtf}

 Klypin et al. (1999) and Moore et al. (1999) noted a 
deficiency in  the number of low mass galaxies (satellites) found within the halo 
of the Milky Way (MW), with respect to the number of subhalos predicted by \LCDM.
Hence the  tag  ``missing satellites problem'' arose. Furthermore, Boylan-Kolchin, Bullock 
\& Kaplinghat (2011) noted that some of the MW subhalos predicted by \LCDM /AM
--- where AM is he Abundance matching technique discussed in section \ref{bardef} 
--- are significantly more massive than any of those observed, hence suggesting the 
paradox that the MW is missing a population of subhalos TBTF: ``too big to fail''.
Were the TBTF phenomenon be noted in the MW alone, its occurrence could be attributed 
to small number statistics. However, more recently Tollerud et al. (2014) have shown 
that M31 also has a TBTF problem; in fact, Ferrero et al. (2012) and Papastergis 
et al. (2015) showed that the population of field dwarfs do so as well. 
The latter is an important result, since the environmental conditions of field 
dwarfs --- which are among the most isolated of galactic systems in the  Universe 
(see section \ref{sec:autocorr})--- are very different
from those within the virial radius of a galaxy as massive as the MW. 
The paucity of low mass galaxies, in comparison with the abundance of low mass halos
predicted by \LCDM, has been known for a long time, appearing in a variety of guises,
e.g. as the  ''void phenomenon'' (Peebles 2001) and in the derivation of luminosity,
HI mass, rotational velocity functions (Baldry et al. 2008;  Klypin et al.2014; Zwaan et al. (2010)
Papastergis et al. 2014 and references therein). The simplest and most direct scenario
for the investigation of the connection between observations and \LCDM ~predictions
is that offered by the rotational velocity function. It presents the least intrusion
by baryonic physics processes, both it and the halo velocity responding to the
radial distribution of mass. Because the HI gas extends radially farther than any
other directly observable component, the observed rotational velocity gets as close
to the halo velocity as it gets. 

\begin{figure}
\begin{center}
  \includegraphics[width=0.80\textwidth]{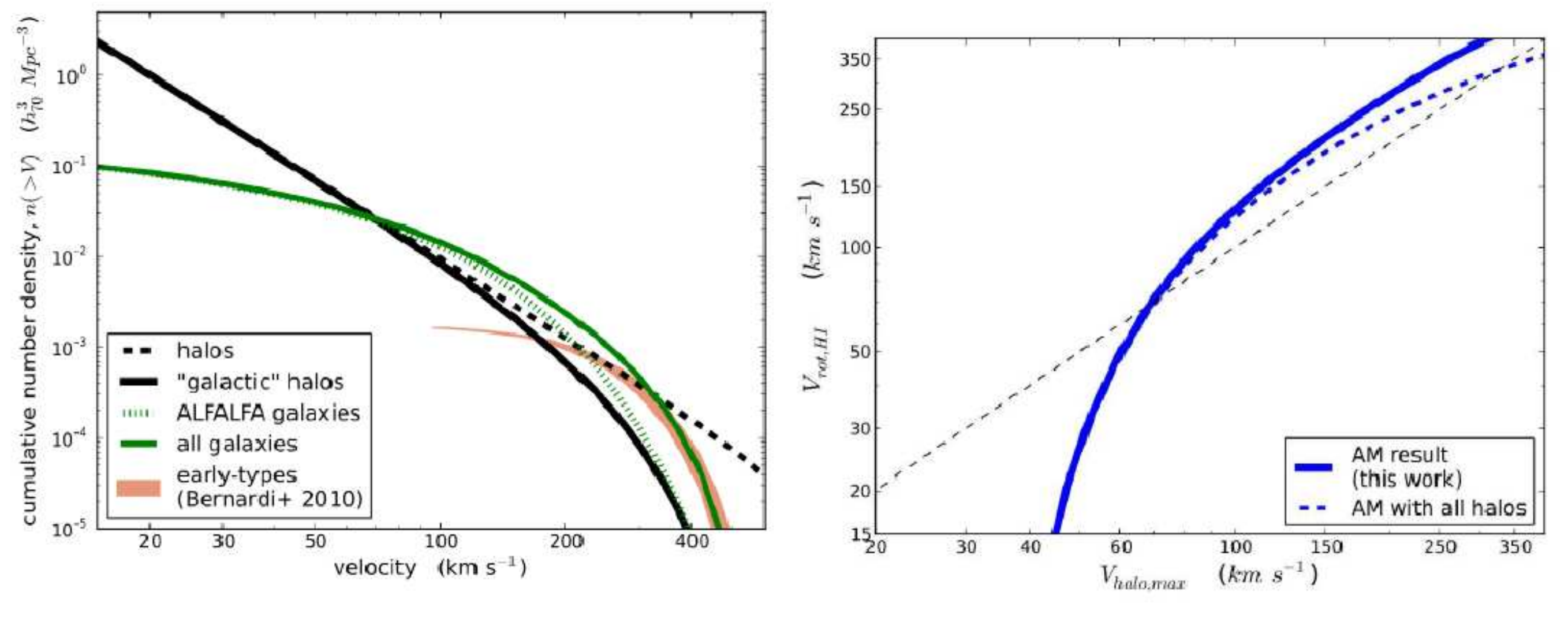}
\caption{\footnotesize {{\bf Left:} The solid black line is the \LCDM ~cumulative
number density of halos as a function of their maximum circular velocity $V_f$, 
as derived from the Bolshoi simulation. The solid green line is the cumulative number density of
ALFALFA galaxies' plotted as a function of their rotational velocity $V_{rot}$.
{\bf Right:} The blue line is the average relation between $V_h$ and $V_{rot}$, as
obtained via AM; the dashed black line is merely for 1-to-1 reference. Ignore the other
lines. Credit: Papastergis et al. (2015).
}}
\label{fig:VelF}      
\end{center}
\end{figure}

The cumulative velocity functions of the
$\alpha.40$ sample of the ALFALFA survey, vis-a-vis the halo Velocity function, are 
shown in the left panel of Figure \ref{fig:VelF}, respectively by the green solid line 
and the black solid line. The average relation $V_h(V_{rot})$, as derived via the AM 
technique, is shown by the blue line in the right panel of the figure. The two 
velocity functions roughly track each other for velocities greater than $\sim 60$ \kms,
but part decisively at velocities lower than that. In the derivation of the velocity
function of the HIPASS survey, Zwaan et al. (2010) reported the mismatch to take
place for velocity widths $< 100$ \kms, although with greater uncertainty than yielded
by ALFALFA' $\alpha.40$, as the latter includes $\simeq 8\times$ as many detections
of low mass sources as the HIPASS catalog (cf. Section \ref{sec:HIMF}). The $V_h(V_{rot})$ 
relation inferred via AM predicts that galaxies with $V_{rot}\sim 15-20$ 
\kms ~reside in halos with $V_h\sim 40-45$ \kms; halos with $V_h<40$ \kms would be 
expected not to host any galaxies at all. 

\begin{figure}
  \includegraphics[width=0.80\textwidth]{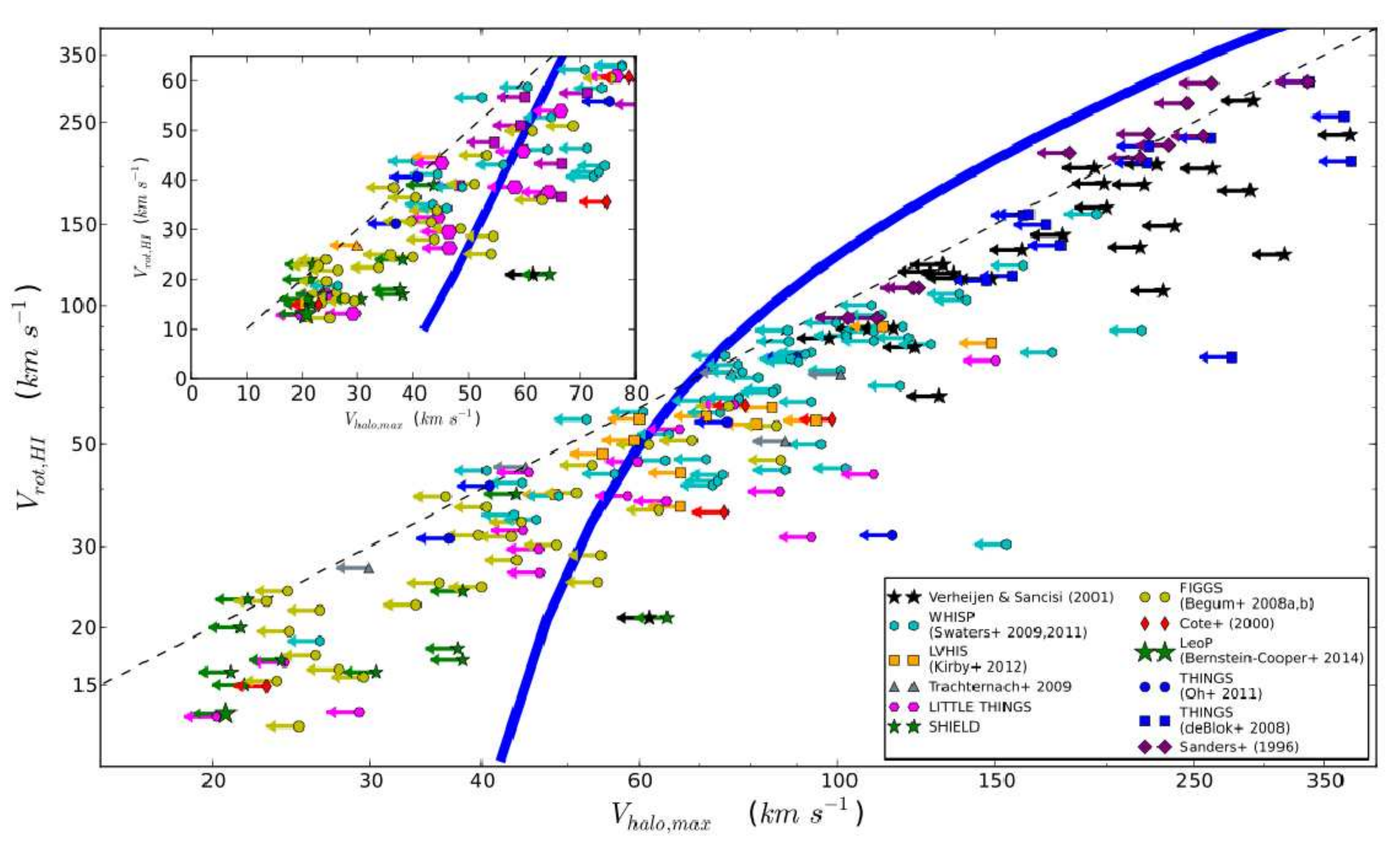}
\caption{\footnotesize {Distribution in a $V_h$ vs $V_{rot}$ of 194 galaxies with resolved, 
interferometric HI rotation curves. The solid blue line is the same function 
as displayed in the right hand panel of Figure \ref{fig:VelF}. The $V_{rot}$ value is the velocity
as given by the last recorded data point of the rotation curve of each galaxy,
after correction for disk inclination and turbulence; the $V_h$ value is the 
maximum velocity of the most massive halo compatible with the galaxy's rotation
curve. The inset panel is a zoom-in, in linear axes, of the low velocity region
of the larger diagram. Credit: Papastergis et al. (2015).
}}

\label{fig:testAM}      
\end{figure}

In an effort to verify the prediction of 
\LCDM/AM, Figure \ref{fig:testAM} shows a preliminary result displaying the distribution 
of 194 galaxies with resolved, interferometric HI rotation curves, in a $V_h$ vs $V_{rot}$
plot (symbols and colors identifying data sources as coded within the plot). These data
allow an independent estimate of the most massive Navarro-Frenk-White (NFW) halo 
compatible with the $V_{rot}$ 
at the last measured point of the rotation curve, i.e. they allow a determination of an upper
limit of $V_{halo,max}$ for each galaxy, and thus to test the \LCDM/AM predictions.
The solid blue line in Figure \ref{fig:testAM} is the same function as displayed in the 
right hand panel of Figure \ref{fig:VelF}, i.e. the locus of the \LCDM ~predicted relation 
$V_h/V_{rot}$, as delivered by AM. The value of $V_{rot}$ for each galaxy is inferred from the
last measured point of the rotation curve, corrected for inclination; the value of $V_h$
(labelled in the Figure as $V_{halo,max}$) is that of the most massive NFW halo compatible
with the $V_{rot}$ at the last measured point of the rotation curve, i.e. each $V_{halo,max}$
is an upper limit. The test, as shown in Figure \ref{fig:testAM}, fails to confirm the
\LCDM ~prediction. The steep downturn predicted by AM, for velocities lower than about 60 \kms
~is not compatible with the observed data. The halos hosting galaxies with $V_{rot} < 30$ \kms 
~appear to have significantly lesser values of $V_h$ than predicted by \LCDM/AM.

The steep slope of the predicted relation $V_h/V_{rot}$ between $V_h \simeq 40$ and 
60 \kms ~raises another issue. Low velocity galaxies with $V_{rot} < 50$ \kms ~ would
``crowd'' over a narrow range of halo velocities. One would thus expect a change in the
slope of the baryonic Tully-Fisher relation, becoming shallower for $V_{rot}<50$ \kms.
That does not seem to be the case, as shown in Figure \ref{fig:LeoP}, 

For the \LCDM/AM approach
to yield predictions more closely compatible with observations in the low end
regime, it would be necessary (i) to either increase the low end slope of the VF
or (ii) to reduce the low end slope of the halo VF. Option (i) would require 
counting galaxies of low $V_{rot}$ which so far have been missed by surveys, 
while option (ii) requires hiding halos by making their baryons invisible.
The burden in correcting the mismatch between the galaxy and halo cumulative 
velocity functions near the low velocity end is a heavy one.
Warm, rather than cold DM particles with mass of few keV, could suppress 
structure on small scales and lower the counts of small halos. This
is a radical approach, which may 
however fail other astrophysical tests (Papastergis et al. 2015). Less radical
solutions may exist. \LCDM ~simulations of high resolution and inclusive 
of the treatment of fully implemented baryonic processes may reveal modifications 
of the properties of low mass halos that would reduce their visibility and thus
reduce small halo counts. In a recent review, Pontzen \& Governato (2014) underscored 
the importance of this solution (see also Governato et al. 2010 and 2012). 
Brooks \& Zolotov (2014) and Sawala et al. (2014) have 
shown that proper account of supernova feedback and tidal stripping of subhalos
by the interaction with the central galaxy's halo can match the observed properties of the 
MW and M31 satellites, and thus explain the TBTF problem in those two systems.
However, some of these processes  depend on the vicinity of a massive 
galaxy, and can explain the TBTF syndrome in MW and M31 subhalos, but should
be less important in the field. It thus remains unclear whether full hydrodynamical 
simulations can save the day for \LCDM ~in low galaxy density environments.

Inspection of Figure \ref{fig:testAM} shows that low mass halos detectable at 21cm, 
albeit not abundant, do exist. Examples of such a population are discussed
in section \ref{dark}:  Leo P was detected by ALFALFA, which motivated a search for
an optical counterpart (Section \ref{uchvc}); like AGC 229385 (Section 
\ref{gallery} it is undergoing a mini-starburst. It is located 1.7 Mpc away; it 
would not have been detected if it were $\simeq 2.5\times$ farther away. Likewise, 
AGC 229385 is also undergoing a mini-starburst; its distance is uncertain, but it appears
to reside within a low mass halo. The same holds for its companion, AGC 229384, which
has no optical counterpart. Although several other objects with similar properties
are detected by ALFALFA, the whole detected lot of ``almost dark''sources is not
sufficient to account for more than a tiny fraction of the mismatch evidenced in 
Figure \ref{fig:VelF}. Could other objects exist, more abundant but just below the 
depth of the current surveys?

\subsection{Too Shy to Shine}
\label{TSTS}

Geha et al. (2012) recently made a discovery of high relevance to the TBTF
issue. Extracting a sample of dwarf galaxies from the SDSS 
spectroscopic data base (objects of stellar mass $M_*$ between $10^7$ and $10^9$ \msun), 
they queried what fraction of that population is currently star forming.  A galaxy was 
classified as star forming if H$\alpha$ emission is detectable and there is evidence 
of a strong 4000\AA ~break. Two subsamples were drawn according to a simple environmental
criterion: the projected distance to the nearest massive galaxy. If that
distance is greater than 1.5 Mpc, the dwarf object is part of the ``field''
population. By these criteria, they found that more than 99\% of 2951
field dwarfs are currently forming stars. 

This finding is surprising, since the history of star formation in dwarf 
galaxies is thought to be episodic (Tolstoy et al. 2009) with a typical duration 
of starburst events in dwarfs of 200 to 400 Myr (McQuinn et al. 2009). 
If the typical field dwarf is star-forming for only a fraction of the time,
a complementary fraction of field dwarfs must be in a quiescent state. If the
average field dwarf has only a few starbursting episodes over a Hubble time,
the uncounted, dormant field dwarfs could account for the missing objects
responsible for the steep drop in the $V_{rot}$ vs. $V_H$ relation in
Figure \ref{fig:testAM}. 

Halos of mass lower than $\sim 10^{10}$ \msun ($V_h\sim 30$ \kms) ~may have been able
to form stars before reionization, but unable to accrete fresh gas from the IGM and
thus to form stars after reionization (Hoeft et al. 2006). However gas accretion
and star sormation may have become possible at later epochs, especially in low density 
environments (Ricotti 2009). In fact some do now. These items of evidence would suggest 
that between star formation episodes field dwarf galaxies can be dark, their optical 
and HI emissions being quenched. The histories of star formation of nearby dwarf
systems would suggest that the fraction of such objects --- unaccounted by surveys
and TSTS (too shy to shine) --- is large. A similar conclusion has been reached by 
Kormendy \& Freeman (2014,2015) in their investigation of the scaling laws of dark 
matter halos in late-type and dwarf spheroidal galaxies.

%
And the skies may be crowded with low mass halos, ready --- like Salpeter \& Hoffman's 
(1995a,b) ``Cheshire Cat'' --- to modestly shine.

\section{The HI Frontier: Tracing HI over Cosmic Time}
\label{future}
While the low redshift surveys provide measures of the fundamental
relations between stars, atomic gas and dark matter in the current galaxy population,
future efforts will aim to explore how the HI content evolves along with
the stellar and molecular components over cosmic time. At the time of
this review, we find the field of HI surveys poised to exploit new
facilities offering huge advances in bandwidth and field of view, leading
to a future revolutionary instrument, the Square Kilometre Array.
Here we discuss the benchmarks that are likely to define the surveys
that will take place in the next years and then by outlining 
representative ones that are on-going or planned as of now.

\subsection{Designing Future HI Surveys: The Telescope Figure of Merit}
\label{FoM}

New facilities are being designed or under construction, which will be used to
carry out ambitious HI surveys in the next decade. In this section we briefly
visit the announced characteristics of the telescopes and an even briefer
one of some of the surveys which will dramatically expand our view of the
HI Universe. A useful parameter often used for the comparison of telescopes 
and surves is a Figure of Merit (FoM)
\begin{equation}
FoM \propto (A_{eff}/T_{sys})^2 \Omega_{fov} BW
\label{eq:FoM}
\end{equation}
where $A_{eff}$ is the telescope aperture efficiency, $T_{sys}$ the system temperature
of the receiver, $\Omega_{fov}$ is the solid angle of the telescope field of view
and $BW$ is its spectral bandwidth. 

Assume arbitrarily a unit FoM
corresponding to an Arecibo-like telescope with an $A_{eff} \simeq 4\times 10^4 ~m^2$,
a $T_{sys}=25$ K and a single pixel field of view. Then the same telescope operating
over the same bandwidth with a 7-feed array and a degraded $T_{sys}$ of 31 K, as would be
the Arecibo telescope with the ALFA focal plane array, would have a $FoM=4.7$. FAST, an 
Arecibo-like telescope currently under construction in southern China, will gain by a factor 3 in 
$A_{eff}^2$, and with a 19 feed planned focal plane array it should yield a FoM$\simeq 39$,
a huge gain over Arecibo. MeerKAT, under construction in South Africa, will be a "traditional"
synthesis array of 64 antennas while Apertiv and ASKAP, respectively operating in Holland
and Australia, will be arrays of between 24 and 36 telescopes, each getting a big kick in
FoM by the installation of focal plane phased arrays (FPPAs) in each of its antennas. While the 
$T_{sys}$ performance of FPPAs is still somewhat uncertain, the multiplexing 
advantage of the FPPA technology is extraordinary and should pay back handsomely. 
Plans at Arecibo and FAST are also afoot for the construction of respectively 40 and 100 beams
FPPAs. At these single dish facilities, optimizing the performance of a single
FPPA by, e.g., cooling the unit, may allow achieving a $T_{sys}$ below 50 K.

Important concerns not contemplated by the FoM as defined above include the impact
of RFI and source confusion, as well as the fact that interferometric arrays will make 
it possible to spatially resolve HI sources out to much larger distances than will AO 
and FAST. 

As discussed in Section \ref{rfi} and above, RFI is a serious impediment
as future surveys explore further into the universe and the redshifted HI
line shifts to lower and lower frequencies where RFI threatens to become a limiting factor.
Interferometers are less prone to the impact of transient or time-varying RFI
as the path length to individual elements of an array varies. However, receiver
saturation will induce bad array data and therefore can have a serious impact
even on interferometric surveys. While some telescope sites are located far from
urban areas in ``radio quiet'' zones, satellite transmissions are increasingly 
rendering bands entirely unusable, regardless of Earth location. Alert monitoring,
identification of RFI sources and legal protection are increasingly important.

Another critical issue for the design of future surveys to measure
possible evolution of the HI mass function is the impact of
source confusion (Duffy et al. 2012; Delhaize et al. 2013;
Jones et al. 2015a, 2015b). While the addition of the velocity dimension
reduces the impact of confusion on spectral line surveys, confusion 
will be particularly critical for future surveys with single dish telescopes
like Arecibo and FAST. Because of their large beams, confusion will start
becoming a problem at moderate redshifts. Source confusion may also constrain 
some of the future deep surveys with synthesis arrays, particularly if full 
angular resolution is not achieved.
In general, the impact of confusion on the derivation of the HI mass function 
and its evolution with $z$ is 
to increase the derived value of the knee-mass M$_{*}$ and to steepen the
faint end slope $\alpha$.
The fort\'{e} of the large aperture single dishes will be the
relatively nearby universe, where their sheer detection sensitivity
will allow the detection of objects to very low HI mass, enabling
studies of the faint end of the HI mass function to log M$_{HI}$/M$_{\odot}
\sim 10^6$. On the other hand, the wide shallow interferometric surveys
WALLABY and WNSHS (see Table \ref{tab:skapathtab}) will be only
negligibly affected by confusion. Confusion will also have an impact on
the potential of survey data bases to be mined with the stacking technique,
as discussed in Section \ref{stack}. Jones et al. (2015b) show that
the stacking potential of DINGO UDEEP will be affected 
if ASKAP is unable to achieve a resolution of 10\arcsec~
for that experiment. In principle, ASKAP should be able to
deliver the higher resolution if sufficient computational
power is available to the DINGO UDEEP program. Due
to their planned angular resolution, CHILES and LADUMA 
(see Table \ref{tab:skapathtab}) should be relatively unaffected
by confusion.

\subsection{Current and Planned Interferometric Surveys}
\label{planned}
In preparation for the deployment of the SKA pathfinders (Arecibo-AO40,
FAST, WSRT/APERTIF, ASKAP and MeerKAT),
a number of pilot programs to search for HI in emission at higher redshift have been
recently undertaken or are underway. Table \ref{tab:highztab} summarizes the
main characteristics of recent and on-going pilot studies which are opening
the HI window into intermediate redshifts while the current plans for
some of the pathfinder surveys are presented in Table \ref{tab:skapathtab}.

\subsubsection{HI Emission beyond the Local Universe: First Results}
\label{highztoday}
Exploiting the huge collecting area of the Arecibo telescope, the pointed HIGHz program
of Catinella \& Cortese (2015) detected HI emission from 39 actively star-forming spirals
at $z > 0.16$. The targets were selected from the SDSS database
on the basis of their optical morphology
(well formed, non-interacting disks) and H$\alpha$ line emission (modest H$\alpha$
equivalent width). As selected, these are relatively isolated, massive gas-rich disks,
all with HI masses of $2-8 \times 10^{10}$ \msun, stellar masses $2-22 \times 10^{10}$ \msun,
and SFRs of 3-35 \msun~yr$^{-1}$. While they have unusually high HI gas fractions and
very blue (NUV-r) colors, their gas content is {\it not} unusual given their UV and optical
properties. This result suggests that the galaxies are the higher redshift analogues of the
most massive and largest disks found in the ALFALFA-selected HIghMass sample by Huang
et al. (2014) and in fact have HI depletion timescales on the order of $\sim$3 Gyr,
typical of those found in normal star forming galaxies locally.
 Despite their large HI masses and the large collecting area of Arecibo,
typical integration times for these detections amounted to 2-4 hours ON-source
(typically less than one half of the total time per target including bandpass subtraction
but not accounting for data rejected because of RFI or standing waves).
The sensitivity required to detect the very weak HI signals from distant
galaxies remains a challenge.

\begin{table}[!t]
\caption{Representative Pilot HI Emission Studies at Higher Redshift}
\smallskip
\begin{center}
\begin{tabular}{ccccc}
\hline
\noalign{\smallskip}
Survey & Selection  & $z$ &  N$_{det}$ & Ref\\
\noalign{\smallskip}
\hline
\noalign{\smallskip}
HIGHz           &  targeted SDSS-selected spirals         &  0.17-0.25  &  39    & $^a$ \\
\quad           &    \quad                  &  \quad      & \quad & \quad \\
GMRT/A 370      &  field of cluster A 370$^\dagger$ &   0.37      & $^\dagger$  & $^b$ \\
GMRT/Cosmos     &  single pointing, zCosmos field$^\dagger$ &  0.345-0.387  &  $^\dagger$ &$^c$ \\
BUDHIES/WSRT    &  fields of clusters A 963 and A2192     &  0.16-0.22  & 150  & $^d$ \\
CHILES-pilot    &  single pointing of COSMOS field$^\dagger$ & $<$0.193  &   33   & $^e$ \\
\noalign{\smallskip}
\hline
\end{tabular}
\end{center}
{\small
$^\dagger$ stacking technique used to derive average properties of optically
selected samples\\
$^a$ Catinella \& Cortese (2015)\\
$^b$ Lah et al. (2009) \\
$^c$ Rhee, J., Briggs, F.H., Lah, P., Chengalur, J.N. (in prep)\\
$^d$ Verheijen et al. (2007);
see update discussed in Jaff\'{e} et al. (2013)\\
$^e$ Fern\'{a}ndez et al (2013)\\
}
\label{tab:highztab}
\end{table}

The SKA pathfinder arrays will be largely dedicated to the conduct of legacy
surveys, including ones devoted to detecting the HI line or targeting the
well-known ``deep field'' for which large multiwavelength datasets enable
the use of the stacking techniques discussed in Section \ref{fig:stack1}.
For several of these programs, no HI was
detected from individual objects, but limits on the aggregate
are set via the stacking of signals at the 3-d positions of optically-known galaxies.
But the detection of individual targets is beginning to pick up. Verheijen et al. (2007)
report the detection of 42 galaxies at $z \sim$ 0.2, in a pilot of the BUDHIES/WSRT
survey; more recently Jaff\'{e} et al. (2013) update that number to 150. The combination
of stacking and individual detection has permitted Verheijen et al. (2007) to suggest
that the blue galaxies in Abell 963 are relatively HI poor compared to similarly
blue disks outside the cluster.
And, as a concept demonstration of the potential for really deep interferometric HI imaging,
the CHILES pilot program (Fernandez et al. 2013) exploited
50 hours with the Very Large Array to search a 34$\times$34 \arcmin~ region of the
COSMOS field at 0$<z<$0.193. The pilot experience yielded detection of 33 galaxies,
30 of which had been detected spectroscopically at optical wavelengths.
These programs have been intended mainly to demonstrate the efficacy of techniques and
predicted outcomes of much more significant investments of telescope time
with telescope arrays, and have successfully made their point.

\subsubsection{HI Absorption Studies}
\label{futHIabs}
An alternative approach to the study of the hydrogen content of distant
objects, mentioned only briefly here, relies on
the detection of HI in absorption,
either in the HI 21cm line in the direction of strong radio continuum sources
or as Lyman-$\alpha$ absorption in the spectra of distant quasars.
Unlike flux-limited HI emission line studies which are 
mass-limited at any distance and reflect both the cold and warm
gas, absorption line detection is column-density
limited (at fixed continuum strength) and therefore distance-independent
and most sensitive to the cold gas in galaxies (Rao \& Briggs 1993).
Absorption line studies probe the HI environment in the
direction of strong radio sources with absorption detected
both in close association with the radio source (``intrinsic absorbers'', e.g.
Roberts 1970; Curran et al. 2013; )
and in absorbers along the line of sight to the
background source (``intervening absorbers'', e.g., Brown \& Roberts 1973;
Roberts et al. 1976; Tanna et al. 2013;
Kanekar 2014). Such measurements,
in combination with multiwavelength studies of other tracers,
provide a powerful probe of the environments of strong radio sources
and, in the small fraction that arise from intervening
absorbers, the cosmic mass density of neutral gas in Damped Lyman $\alpha$
(DLA) at different redshifts. Nearly all measurements of the latter
come from optical observations of the redshifted Ly$\alpha$
absorption along lines of sight to background QSOs (e.g. Wolfe et al. 1986,
Prochaska \& Wolfe 2009, Noterdaeme et al. 2012).

However, additions to new bandwidth capabilities of the
GMRT and JVLA and the wide area coverage of new surveys is enabling
a new generation of HI absorption line studies. Samples of HI synthesis
mapping of intrinsic absorbers are growing (e.g. Chandola, Gupta \&
Saikia 2013; Gereb et al 2015a,b). The synthesis maps yield both
the structure of the underlying continuum source and the kinematics of
the gas. In some cases, the signature of gas infall onto
the active nucleus (e.g., Srianand et al. 2015) is evident, while in others, outflows
are more likely (Gereb et al. 2015b). Future systematic studies will
explore the causal relationships between HI absorption and the impact
of mergers and feedback.

Making use of an early version of the
ALFALFA dataset, Darling et al. (2011) discuss
the potential to use the future wideband capabilities of the
SKA pathfinders to look for the HI analogues of DLA absorbers.
A first demonstration of the power of these future surveys
using the Australian SKA Pathfinder (ASKAP) searched the
full redshift range 0.4 $< z <$ 1.0, yielding a new detection
at $z = 0.44$ (Allison et al. 2015)
in anticipation of the proposed ASKAP/FLASH survey. Over the next years,
FLASH will use the 300 MHz capability of ASKAP to search for HI absorbers
both in the radio source hosts and in intervening absorbers,
starting as ASKAP early science with targeted fields. The 
MeerKAT Absorption Line Survey (MALS) will survey the entire bandwidth
from 0.58 to 1.75 GHz
of MeerKAT to search for both HI and OH absorbers out to z=1.8.
The resultant
dataset will provide uniquely for investigations of the
evolution of the HI mass density at intermediate redshift.

\begin{table}[!h]
\caption{Representative Planned SKA Pathfinder Surveys}
\smallskip
\begin{center}
\begin{tabular}{lrrcccl}
\hline
\noalign{\smallskip}
Survey & Res.$^\#$ & Area   &  $z$ & N$_{det}^\dagger$ & Ref & Note\\
       & \arcsec    & \sqd   &      &             &     &  \\
\noalign{\smallskip}
\hline
\noalign{\smallskip}
\bf{VLA-B}        &       &           &               &        &      & \\
\quad CHILES      &  5    &  0.8      &  $<$0.5       &  300   & $^a$ & COSMOS deep field\\
\bf{WSRT/APERTIF} &       &           &               &        &      &  \\
\quad WNSHS$^*$       & 15    & 3500$^*$&  $<$0.26      & 50000$^*$  & $^b$ & Shallow, wide area   \\
\quad MediumDeep$^*$  & 15    & 200$^*$ &  $<$0.26      & 1$\times$10$^5$$^*$  & $^c$ & Selected fields  \\
\bf{ASKAP}        &       &           &               &                  &      & \\
\quad WALLABY     & 30    & 30000     &  $<$0.26      & $>3\times$10$^5$   & $^d$ & Shallow, wide area \\
\quad DINGO-DEEP  & 10    & 150       &  0.1--0.26    & 50000  & $^e$ & GAMA region    \\
\quad DINGO-UDEEP & 10    &  60       &  0.1--0.43    & 50000  & $^e$ & GAMA region    \\
\quad FLASH       & 30    & targeted  &  0.5-1.0         &  few 100s        & $^f$ & HI absorption   \\
\bf{MeerKAT}      &       &           &               &                  &      &  \\
\quad MHONGOOSE$^\&$    & 12    & 30 x 0.8  &         &    30$^\&$       & $^g$ & 30 nearby galaxies   \\
\quad LADUMA      & 12    &   4       &   $<$1.4      &  10000           & $^h$ & ECDF-S deep field   \\
\quad MALS        & 12    &           &   $<$1.8      &   600            & $^i$ & HI and OH absorbers\\
\noalign{\smallskip}
\hline
\end{tabular}
\end{center}
{\small
$^\#$ Expected image resolution. For some surveys, the resolution may be practically limited by
the heavy demands placed on computational resources.\\
$^\dagger$ Number of predicted HI detections. In addition,
stacking will also be important for many surveys.\\
$^*$ Survey area not yet specified, so numbers only approximate\\
$^\&$ Focused on nearby galaxies but background surveyed commensally.\\
$^a$ http://chiles.astro.columbia.edu\\
$^b$ http://www.astron.nl/phiscc2014/Documents/Surveys/jozsa\_dwingeloo\_wnshs.pdf\\
$^c$ http://www.astron.nl/phiscc2014/Documents/HI\_in\_Galaxies/Verheijen\_PHISCC2014.pdf\\
$^d$ http://www.atnf.csiro.au/research/WALLABY\\
$^e$ http://internal.physics.uwa.edu.au/$\sim$mmeyer/dingo/welcome.html\\
$^f$ http://www.caastro.org/research/evolving/flash\\
$^g$ http://mhongoose.astron.nl\\
$^h$ http://www.ast.uct.ac.za/laduma\\
$^i$ http://www.eso.org/sci/meetings/2011/gas2011/Talks/Wednesday/Gupta$\_$chile2011.pdf\\
}
\label{tab:skapathtab}
\end{table}

\subsubsection{Exploring HI across Cosmic Time: The Next Five Years}
\label{futHIems}
The present review finds
itself on the embarkation of several major surveys, including the full CHILES
deep survey with the VLA and others planned to begin in the next few years using
WSRT/APERTIF, ASKAP and MeerKAT. Table \ref{tab:skapathtab} summarizes some of the
representative SKA pathfinder surveys as they are currently envisage. It should be noted
that the final configurations and capabilities of the pathfinders are not yet set;
the parameters listed in Table \ref{tab:skapathtab} are current goals and may prove
to be optimistic. However, all of them are likely to produce results in the next
few years, taking advantages of the new capabilities now being deployed at these
facilities, with first early science results already illustrating their scientific
promise (e.g. Lucero et al. 2015;  Hess et al. 2015; Serra et al. 2015).
Of particular note, the several
planned deep surveys, in combination with major efforts at other
wavelengths, will allow a first exploration of the evolution of the
HI mass function over the the last few billion years, laying the groundwork
for the next generation of surveys exploiting the square kilometer of
collecting area promised by the full SKA.

Beyond the sheer numbers of detections, important studies of the HI morphologies and kinematics
for thousands of galaxies will be enabled by these future surveys. Among the pathfinders,
the angular resolution of ASKAP ($\sim$30\arcsec) and WSRT/APERTIF ($\sim$15\arcsec) will allow
WALLABY and WNSHS to obtain maps with more than 10 beams across for
800 and 300 galaxies respectively, and more than 35,000 and 10,000 galaxies will
be sampled over 3-10 beams, respectively. As a follow-on to the WSRT HALOGAS
survey probing to low column densities, (Heald et al. 2011),
the MHONGOOSE program will spend 200 hours per object staring at 30 nearby
galaxies, delivering maps of unprecedented column density depth and probing the processes by which
gas supplies are both depleted and replenished in spiral disks.

\section{The Future of HI Surveys}

At the time of writing of this review, plans for SKA-mid, intended to be operational
in the early 2020's, suggest that some 1500 hours per year
should be available for HI-related studies. A unique feature of the
heterodyne techniques used in cm-wave astronomy is the potential for commensality,
whereby the same acquired datastreams 
can be used simultaneously for both continuum and spectral line surveys (i.e., the data 
are acquired in spectral line mode and the continuum experiments sum the spectral channels).
As has been the case for the execution
of the ALFA surveys at Arecibo, there is very significant potential
for commensality between future continuum and HI line surveys so that the amount of
SKA-mid telescope time
available for HI science may be even larger. On-going activities of the
SKA Pathfinders HI Science Coordination Committee (PHISCC) revolve around coordinating
the planned surveys and sharing efforts to maximize scientific productivity (Oosterloo 2015).
These pathfinder surveys will only begin the exploration of the HI frontier at
high redshift, allowing some quantitative input on the dominant form of gas -- and even
baryons -- in the low mass, star forming population.

The present review has focused on the outcome and plans for surveys of the HI content
of individual galaxies or of aggregates of galaxies where stacking is employed.
Other compelling science is enabled by 21 cm HI line studies well beyond the limited
scope of this review. Most notable among them will be: the detection of 
the redshifted HI line from before and during the epoch of reionization
(e.g. Paciga et al. 2013; Yatawatta et al. 2013; Dillon et al. 2014; Jacobs et al. 2015); 
intensity mapping to probe the evolution of the baryon acoustic scale length
with a single probe over a wide range of redshift (e.g., Chang et al. 2008, 2010); 
and using HI measures of
the rotational velocity in applications of the Tully-Fisher relation to
derive the local peculiar velocity field (e.g. Koda et al. 2014;
Hoffman et al. 2015). For many of these and further discussions,
the reader is referred to the science documents associated with the SKA science
case (e.g., Carilli \& Rawlings 2004, a new version of which is in
the process of being updated for which many white papers are available on
the arXiv). Ultimately, the construction of the full SKA
will enable the detection of billions of HI galaxies throughout the 
universe adding a vital ingredient to our understanding of how
gas has been converted into stars throughout the history of the universe.

Poised as we are on a verge of a rising wave
of advanced antenna, electronic, digital, computational and algorithmic technology, the
future of HI survey science will be bright indeed!

\begin{acknowledgements}
We thank the entire ALFALFA survey team for their many contributions over the
last decade (and more) towards
observing, data processing and analysis that have led to a rich ALFALFA harvest.
The ALFALFA survey team at Cornell has been supported by US NSF grants AST-0607007 and 
AST-1107390 to RG and MPH and by continuing support from the Brinson Foundation. 
\end{acknowledgements}

\end{document}